\documentclass[onecolumn]{aa}
\usepackage{txfonts}
\usepackage{natbib}
\usepackage{graphicx}
\begin{document}
\title{\textbf{The puzzling case of GRB~990123: prompt emission and broad-band afterglow modeling}}
\author{A.~Corsi\inst{1,2}\and L.~Piro\inst{1}\and E.~Kuulkers\inst{3}\and L.~Amati\inst{4}\and L.A.~Antonelli\inst{5}\and E.~Costa\inst{1}\and M.~Feroci\inst{1}\and F.~Frontera\inst{4,6}\and C.~Guidorzi\inst{6}\and J.~Heise\inst{7}\and J.~in~'t~Zand\inst{7}\and E. Maiorano\inst{4,8}\and E. Montanari\inst{6}\and L.~Nicastro\inst{9}\and E.~Pian\inst{10}\and P.~Soffitta\inst{1}.}
\institute{IASF-CNR, Via Fosso del Cavaliere 100, I-00133 Rome, Italy.
\and University ``La Sapienza'', Piazzale Aldo Moro 5, I-00185 Rome, Italy.
\and ESA-ESTEC, Science Operations \& Data Systems Division, SCI-SDG, Keplerlaan 1, 2201 AZ Noordwijk, The Netherlands.
\and IASF-CNR, Via Gobetti 101, I-40129 Bologna, Italy.
\and Rome Astronomical Observatory, Via di Frascati 33, I-00044 Rome, Italy.
\and Physics Department, University of Ferrara, Via Paradiso 11, I-00044 Rome, Italy.
\and Space Research Organization Netherlands, Sorbonnelaan 2, 3584 CA Utrecht, The Netherlands.
\and Astronomy Department, University of Bologna, Via Ranzani1, I-40126 Bologna, Italy.
\and IASF-CNR, Via Ugo la Malfa 153, I-90146 Palermo, Italy.
\and INAF, Osservatorio Astronomico di Trieste, Via G.B. Tiepolo, 11 - I-34131 Trieste, Italy.}
\offprints{A. Corsi -- Alessandra.Corsi@rm.iasf.cnr.it}
\date{}
\abstract{
We report on \textit{Beppo}SAX simultaneous X- and $\gamma$-ray observations of the bright $\gamma$-ray burst (GRB) 990123. We present the broad-band spectrum of the prompt emission, including optical, X- and $\gamma$-rays, confirming the suggestion that the emission mechanisms at low and high frequencies must have different physical origins. In the framework of the standard fireball model, we discuss the X-ray afterglow observed by the Narrow Field Instruments (NFIs) on board \textit{Beppo}SAX and its hard X-ray emission up to $60$~keV several hours after the burst, detected for about $20$ ks by the Phoswich Detection System (PDS). Considering the $2-10$ keV and optical light curves, the $0.1-60$ keV spectrum during the $20$ ks in which the PDS signal was present and the $8.46$~GHz upper limits, we find that the multi-wavelength observations cannot be readily accommodated by basic afterglow models. While the temporal and spectral behavior of the optical afterglow is possibly explained by a synchrotron cooling frequency between the optical and the X-ray energy band during the NFIs observations, in X-rays this assumption only accounts for the slope of the $2-10$~keV light curve, but not for the flatness of the $0.1-60$~keV spectrum. Including the contribution of Inverse Compton (IC) scattering, we solve the problem of the flat X-ray spectrum and justify the hard X-ray emission; we also suggest that the lack of a significant detection of $15-60$ keV emission in the following $75$ ks and last $70$ ks spectra, should be related to poorer statistics rather than to an important suppression of IC contribution. However, considering also the radio band data, we find the $8.46$~GHz upper limits violated. On the other hand, leaving unchanged the emission mechanism requires modifying the hydrodynamics by invoking an ambient medium whose density rises rapidly with radius and by having the shock losing energy. Thus we are left with an open puzzle which requires further inspection.
}
\authorrunning{A. Corsi, L. Piro, E. Kuulkers et al.}
\titlerunning{The puzzling case of GRB 990123}
\maketitle
\section{Introduction}
GRB~990123 was one of the brightest $\gamma$-ray bursts detected by the \textit{Beppo}SAX satellite; it was also observed by the Burst and Transient Source Experiment (BATSE) on-board the Compton Gamma-Ray Observatory (CGRO) as trigger \# 7343 \citep{Briggs1999}. The Gamma-Ray Burst Monitor (GRBM) was triggered by GRB~990123 on 1999 January 23.40780 UT, approximately $18$~s after the CGRO trigger. The burst was simultaneously detected near the center of the field of view in Wide Field Camera (WFC) unit 1 \citep{Feroci1999}; the WFC ``quick look'' localized the burst within a radius of $2$~arcmin (99\% confidence level). 

With a redshift of $z=1.6004$ and a luminosity distance of $3.7\times10^{28}$~cm \citep{Kulkarni1999a}, a $\gamma$-ray ($40-700$~keV) fluence of $F_{\gamma}=(1.9\pm0.2)\times10^{-4}$~ergs~cm$^{-2}$ implies an isotropic energy release in $\gamma$-rays alone of about $1.2\times10^{54}$~ergs \citep[see also][]{Briggs1999, Kulkarni1999a}. GRB 990123 would have been notable even just for this reason. Furthermore, it was the first burst from which simultaneous $\gamma$-ray, X-ray and optical emission was detected. The prompt announcement of the burst position resulted in intensive multi-wavelength follow-up observations that brought a wealth of new results: the discovery of prompt optical emission \citep{Akerlof1999}, the detection of short lived radio emission \citep{Kulkarni1999b}, the first observation of a clear break in the optical afterglow light curve \citep{Kulkarni1999a,Fruchter1999,Castro-Tirado1999,Holland2000}, the first constraint on the polarization of a GRB afterglow \citep{Hjorth1999}.

We refer the reader to the paper of \citet{Maiorano2004} for the
complete analysis of the multi-wavelength afterglow data. In this paper we present the \textit{Beppo}SAX GRBM and WFC observations of the $\gamma$- and X-ray  prompt event of  GRB~990123  (section \ref{The burst event}) and analyze the spectral properties of its X-ray afterglow (section \ref{The X-ray afterglow}). We discuss the observed  prompt and afterglow emission comparing them with the predictions of the ``forward plus reverse shock'' standard model. While the broad-band spectrum of the burst confirms a reverse shock origin of the optical flash (section \ref{prompt_discussion}), the multi-wavelength afterglow cannot be readily explained by basic models (section \ref{subsection:afterglow}). Errors on the parameters will be given at $90\%$ confidence level ($\Delta\chi^{2}=2.7$ for a one parameter fit).
\section{The observations}
\subsection{The prompt event}
\label{The burst event}
The burst was detected by the \textit{Beppo}SAX GRBM and WFC. The GRBM \citep{Amati1997,Feroci1997} consists of the 4 anti-coincidence shields of the \textit{Phoswich Detection System}, PDS \citep{Frontera1997,Costa1998} and it operates in the $40-700$~keV energy band. The normal directions of two GRBM shields are co-aligned with the viewing direction of the WFCs. The WFCs \citep{Jager1997} consist of two identical coded aperture cameras, each with a field of view of  40$^{\circ}\times40^{\circ}$ full-width to zero response and an angular resolution of about $5'$. The bandpass is $2$~keV to $28$~keV. The spectral resolution is approximately constant over the bandpass at $20\%$ FWHM.
\begin{figure*}
\centering
\includegraphics[angle=-90,width=\columnwidth]{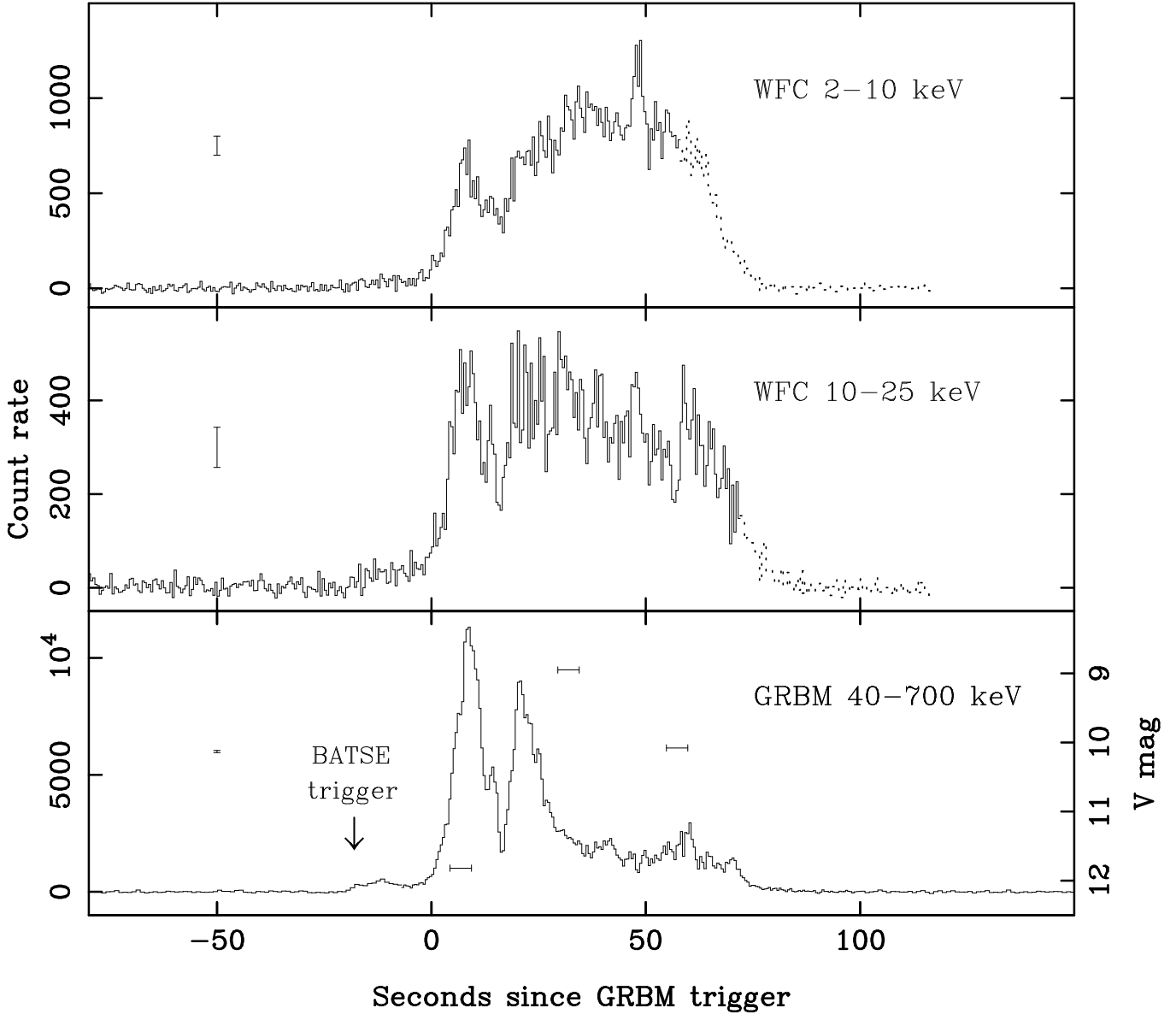}
\caption{Prompt burst profile of GRB~990123 at X-ray 
(WFC: 2--10~keV and 10--25~keV)
and at $\gamma$-ray (GRBM: 40--700~keV) energies. Count rate (counts s$^{-1}$)
is given as a function of time after the GRBM trigger,
i.e.\ 1999, Jan 23, 09:47:14 UT. Since the X-ray burst was close to the
Earth horizon, atmospheric absorption plays an important role.  
At 80~s after trigger the Earth-atmospheric absorption is about
30\%\ at 5~keV and the subsequent decay is partially due to the atmosphere.
The dotted line refers to the part where we loose more than 10\%\ of the 
intensity due to this effect. The two X-ray light curves end when GRB~990123 sets below the Earth horizon.
Note that the $\gamma$-ray light curve is not influenced by the Earth atmosphere. Typical error bars are given in the left part of the panels.
Indicated in the bottom panel are the time of the BATSE
trigger and the prompt optical measurements 
\citep[ROTSE, ][]{Akerlof1999}.}
\label{fig:prompt}
\end{figure*}
In Fig.~\ref{fig:prompt} the time profile of the burst is shown in various band-passes. The $\gamma$-ray signal is detected for about 100~s with the GRBM (Fig. 1, bottom
panel); it presents two major pulses, the brighter of which has a peak intensity of $(1.7\pm0.5)\times10^{-5}$~ergs~cm$^{-2}$~s$^{-1}$ between $40-700$~keV. The total fluence between $40$~keV and $700$~keV is $(1.9\pm0.2)\times10^{-4}$~ergs~cm$^{-2}$; this value improves the one given in \citet{Amati2002}.

The burst profile at lower energies ($2-10$~keV and $10-25$~keV) as measured by the WFC is shown in the top panels of Fig.~\ref{fig:prompt}. Near the end of the measurement the WFC was pointing close to the Earth horizon. Since the atmospheric absorption then plays an important role, the low-energy X-ray light curve was particularly affected: at $80$~s after the trigger the Earth-atmospheric absorption is about $30$\% at $5$~keV and the subsequent decay in the X-ray curve is partially due to the atmosphere. The portion of the light curve where more than 10\%  of the intensity is lost due to this effect is indicated by a dotted curve. The two X-ray light curves end when GRB 990123 sets below the Earth horizon. Note that the GRBM light curve is not influenced by the Earth atmosphere. The structure in the first $80$~s of the X-ray light curves indicates a softening of each pulse, together with an increase in duration with decreasing energy. As remarked by \citet{Briggs1999} and by \citet{Frontera2004}, during the first two pulses the hardness of the GRB is correlated with the intensity. After the second pulse, a hard-to-soft evolution, typical of most GRBs, is seen. In the $2-25$~keV and $2-10$ keV energy bands, the fluence is $(7.8\pm0.4)\times10^{-6}$~ergs~cm$^{-2}$ and $(3.1\pm0.3)\times10^{-6}$~ergs~cm$^{-2}$, respectively; these estimates must be considered as a lower limits due to the eclipse discussed above.

The model fitting of the data is done using the standard forward-folding technique \citep{Briggs1996}: in each instrument we assume a photon model and convolve it through a detector model to obtain a model count spectrum. The model count spectrum is compared with the observed count spectrum and the photon model parameters are optimized so to minimize a $\chi^{2}$ statistic. The $\nu f_{\nu}$ data points are then calculated by scaling the observed count rate in a given channel by the ratio of the photon to count model rate for that channel; this ratio, and therefore the photon data points, are model dependent. 

The spectrum of the burst can be modeled by using the smoothly broken power-law proposed by \citet{Band1993}, whose parameters are the low energy index $\eta_{1}$, the break energy $E_{0}$, the high energy index $\eta_{2}$. If we express $E$ and $E_{0}$ in keV, the Band function has the following expression:
\begin{equation}
N(E)\propto\left(\frac{E}{100~\rm keV}\right)^{\eta_{1}}\times\exp\left(-\frac{E}{E_{0}}\right)
\end{equation}
for $E\le(\eta_{1}-\eta_{2})\times E_{0}$, and: 
\begin{equation}
N(E)\propto\left[\frac{\left(\eta_{1}-\eta_{2}\right)\times E_{0}}{100~\rm keV}\right]^{\eta_{1}-\eta_{2}}\times\exp(\eta_{2}-\eta_{1})\times\left(\frac{E}{100~\rm keV}\right)^{\eta_{2}}
\end{equation}
for $E\geq (\eta_{1}-\eta_{2})\times E_{0}$.
In the case of GRB~990123, the best fit values for the $2-700$~keV integrated spectrum of the burst are: $\eta_{1}=-0.89\pm0.08$, $\eta_{2}=-2.45\pm0.97$ and $E_{0}=1828\pm84$~keV \citep{Amati2002}, where $E_{0}$ is expressed in the GRB cosmological rest-frame; at a redshift of $z=1.6$, it corresponds to an observed value of $E_{0}=703\pm32$ keV.

\begin{figure*}
\centering
\includegraphics[angle=-90,width=\columnwidth]{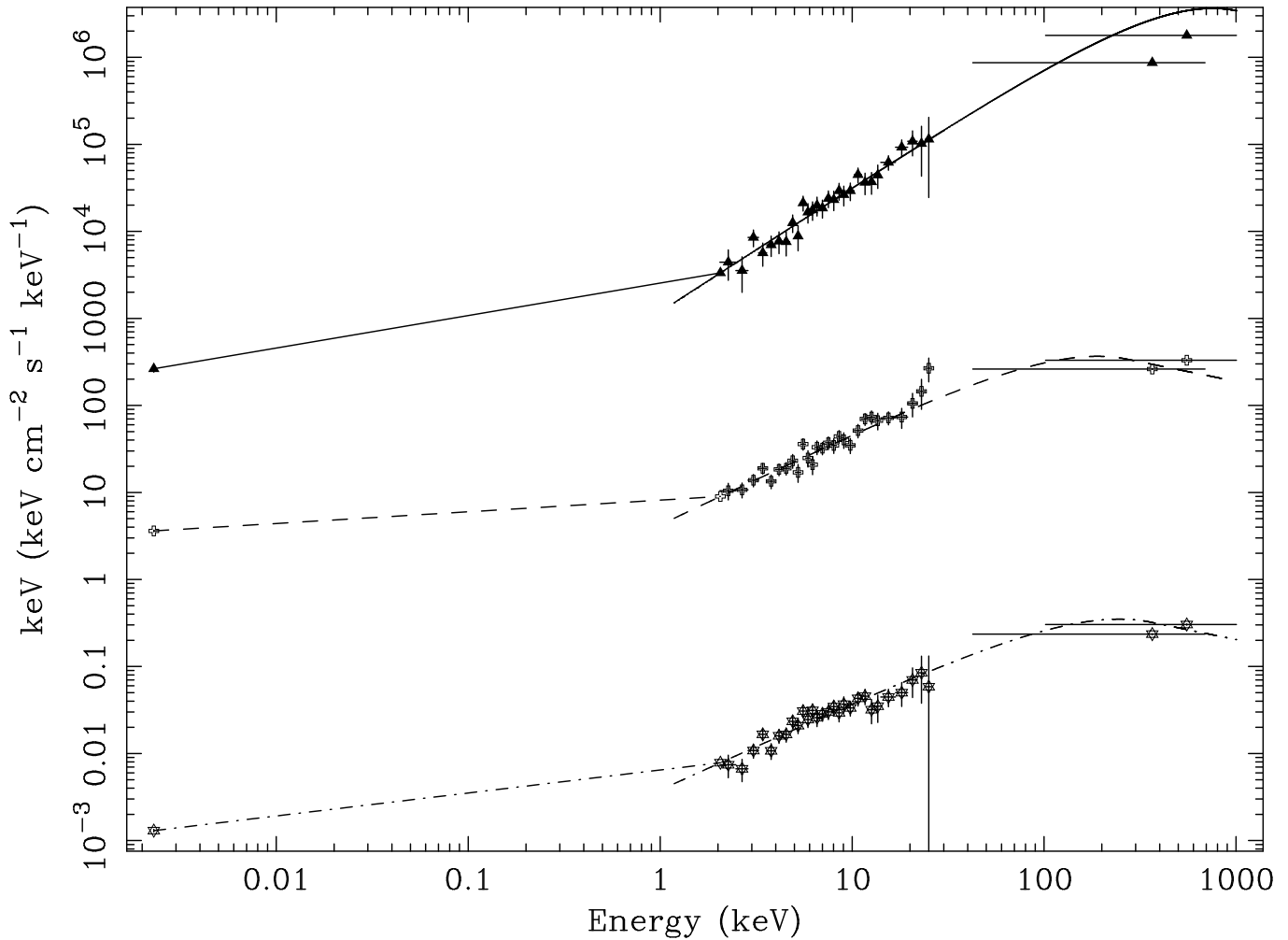}
\caption{Simultaneous multi-wavelength spectra derived at three times during the burst (ROTSE V band, X-ray and $\gamma$-ray, see also Table~1).
The ROTSE data points have been connected with a line to the best fit model at $2$ keV to guide the eye. The solid line connects the points relative to the first ROTSE observation, the dashed line those relative to the second ROTSE observation and the dashed-dotted line those relative to the third one. The corresponding times of the spectra are given in Table 1 in seconds after the GRBM trigger. The data relative to the first (third) time interval have been shifted up (down) of a factor $1000$.}
\label{fig:punti_spettrali}
\end{figure*}

Using the Band function as the model spectrum, we performed a spectral fitting of the WFC and GRBM data at the epochs of the detection of the prompt optical emission \citep{Akerlof1999}, indicated in the lower panel of Fig.~1. In Fig.~\ref{fig:punti_spettrali} we plot the time resolved X- to $\gamma$-ray spectra and the three ROTSE data points; in Table 1 we report the fluxes and the best fit spectral indices in the WFC and GRBM. 
\begin{table*}
\begin{minipage}[t]{\textwidth}
\begin{center}
\caption{Fluxes and photon indices of GRB~990123 prompt emission measured during the first three ROTSE exposures 
\citep{Akerlof1999}.}
\renewcommand{\footnoterule}{}
\begin{tabular}{c c c c c c c}
\hline\hline
\multicolumn{7}{c}{X-ray and $\gamma$-ray data simultaneously with ROTSE} \\
\hline
start\footnote{{Start time since GRBM trigger.}}  &
end\footnote{end time since GRBM trigger.}    &
m$_{\rm v}$             &
Photon index& $E_{0}$ &  2--10 keV flux & 40--700 keV flux \\
%
s &
s &   &
$\eta_{1}$  &  keV& 10$^{-8}$\,ergs cm$^{-2}$ s$^{-1}$   & 10$^{-6}$\,ergs cm$^{-2}$ s$^{-1}$ \\
%
\hline
4.38                    & 9.38                   & $11.70\pm0.07$  &
$-0.57\pm0.06$         & $530^{+1600}_{-230}$&$3.2$& $7.3$ \\
%
29.58                    & 34.58                     & $8.86\pm0.02$   &
$-0.95\pm 0.09$        &  $180^{+90}_{-40}$ &
$5.7$   & $1.3$    \\

54.87                    & 59.87                     & $9.97\pm0.03$  &
$-1.0\pm0.1$         & $250^{+160}_{-70}$ &
$4.8$    & $1.2$ \\
\hline
\end{tabular}
\end{center}
\end{minipage}
\end{table*}

Since the value of $E_{0}$ derived from the $2-700$~keV integrated spectrum is consistent with being above the energy range of the GRBM channels, we set the high energy spectral index $\eta_{2}$ equal to its best fit value for the mean spectrum, $\eta_{2}=-2.45$ \citep{Amati2002}; the values we find for $E_{0}$ in the three time intervals are reported in Table 1.
As can be seen in Fig.~\ref{fig:punti_spettrali}, the spectrum of the burst shows a hard to soft evolution; the spectral photon index for the first ROTSE exposure is $\eta_{1}=-0.57\pm0.06$, harder than the subsequent values of $\eta_{1}=-0.95\pm0.09$ and $\eta_{1}=-1.0\pm0.1$, for the second and last time intervals, respectively.

We find no evidence for an X-ray excess during the first ROTSE observation, as suggested by \citet{Briggs1999}. In fact, our value for the low-energy spectral index is consistent with $\eta_{1}=-0.63\pm0.02$, found by \citet{Briggs1999} analyzing the $10$~keV to $10$~MeV spectrum \citep[see Fig. 3 of ][]{Briggs1999}: from the presence of an X-ray excess below $10$~keV, we would expect to find $\eta_{1}<-0.63$ for $\nu<10$~keV, in contrast with our results. Moreover, performing a spectral fit by using a simple power-law and considering only the WFC data at energies below $10$~keV, yields a best fit photon index of $-0.44\pm0.28$, consistent with the spectral index we find for the overall spectrum.

\subsection{The X-ray afterglow.}\label{The X-ray afterglow}
The \textit{Beppo}SAX follow-up observation lasted from 1999 January $23.65$~UT until January $24.75$~UT.
A previously unknown, bright X-ray source designated as 1SAX J1525.5+4446 \citep{Heise1999}, was detected by the LECS and MECS units, at right ascension $\alpha=15$~h~$25$~m~$31$~s and declination $\delta=+$44$^{\circ}46'.3$ (equinox 2000), with an error-circle radius of $50''$. Within $22''$ of this position, \citet{Odewahn1999} detected a fading optical transient (OT) at magnitude R=$18.2$. During the first $10$~min, the $2-10$~keV photon rate of 1SAX J1525.5+4446 in the MECS is $(0.16\pm0.02)$~count~s$^{-1}$, corresponding to $(1.38\pm0.14)\times10^{-11}$~erg~cm$^{-2}$~s$^{-1}$.

The $2-10$~keV flux steadily decreases with time following a power law decay $F_{\rm X}(t)=F_{\rm X}(6$~hr$)\times(t/6$~hr$)^{\alpha_{\rm X}}$~$\mu$Jy, with $\alpha_{\rm X}=-1.46\pm0.04$ \citep{Maiorano2004}. Here $t$ is the time measured with respect to the $\gamma$-ray event. We refer to \citet{Maiorano2004} for a complete description of the WFC, MECS and PDS data extraction and analysis relative to GRB~990123 light curves in the $2-10$ keV and $15-28$ keV energy bands.

The observed fluence during the observations was $3.1 \times 10^{-7}$~ergs~cm$^{-2}$ ($2-10$~keV) and $7.5 \times 10^{-7}$ ergs~cm$^{-2}$ ($2-60$~keV). Extrapolating to long times after the trigger (about $1000$~s) this implies an X-ray afterglow energy fraction of at least $\sim$11\% that of the prompt X-ray emission ($2-25$~keV).

In the first $20$~ks of the observation, between 1999 January 23.6498 UT and 1999 January 23.8813 UT, the total measured photon rate in the PDS is $(0.20\pm0.05)$~count~s~$^{-1}$, equivalent to $(1.9\pm0.5)\times10^{-11}$~ergs~cm$^{-2}$~s$^{-1}$ ($15-60$~keV). The possible contamination by an X-ray source located in the MECS field has been analyzed by \citet{Maiorano2004}. While its presence affects the PDS data in the second and last part of the observation, it is comparatively negligible during the first $20$~ks: for the first time, an afterglow emission was detected at energies as high as $60$~keV. In the following $75$~ks and the last $70$~ks of the observation, the high energy tail of the afterglow emission ($15-60$~keV energy band) was not detected.

Since the first $20$~ks of the observation are those in which we have the greater statistics and where the PDS signal is present, we focus our attention on the spectrum of the burst relative to this first time interval.
\begin{table*}
\begin{minipage}[t]{\textwidth}
\begin{center}
\caption{Fit results for GRB~990123 X-ray afterglow. $\beta_{\rm X}$ and $\beta_{\rm X}^{\rm IC}$ are the spectral energy indices for the synchrotron and the IC component, respectively (see equation (\ref{eq:s+Ic})); the galactic hydrogen column density $\rm N_{\rm H}$ is fixed to its value at the burst position, $\rm N_{\rm H}=2.1\times10^{20}$~cm$^{-2}$; $\rm N^{\rm z}_{\rm H}$ is the hydrogen column density locally to the GRB site (in units of $10^{22}$~cm$^{-2}$); $\chi^{2}$ is reduced for the degrees of freedom dof. Case [1] refers to a single power-law model spectrum with variable spectral index; case [2] refers to the same model spectrum of case [1], but with a fixed spectral index $\beta_{\rm X}=-1.23$ (see Fig.~\ref{fig:Fnu_indice_fissato}); case [3] refers to a single power-law model spectrum with the fit performed only on the LECS and MECS data points (PDS data excluded); case [4] is the synchrotron plus IC model spectrum, with the spectral index of the synchrotron component fixed at $\beta_{\rm X}=-1.23$ and the one of the IC component at $\beta_{\rm X}^{\rm IC}=-0.73$ (see Fig.~\ref{fig:Fnu_syn+IC} and Fig.~\ref{fig:nuFnu_syn+IC}). $F^{\rm syn}_{0}$ and $F_{0}^{\rm IC}$ are given at $1$ keV, in units of $10^{-3}$~photons~cm$^{-2}$~s$^{-1}$~keV$^{-1}$, in [1], [2], [3] and at $2$ keV, in units of $\mu$Jy, in [4].}
\renewcommand{\footnoterule}{}
\begin{tabular}{c c c c c c c c}
\hline\hline
\multicolumn{8}{c}{First $20$~ks of the NFIs observation}\\
&$\beta_{\rm X}$&$F^{\rm syn}_{0}$&$\beta^{\rm IC}_{\rm X}$&$F_{0}^{\rm IC}$&$\rm N^{\rm z}_{\rm H}$&$\chi^{2}$&dof\\
\hline
$[1]$ &$-0.82\pm0.10$&$2.3\pm0.3$&0&0&$0.42^{+0.88}_{-0.24}$&$1.3$&$26$\\
$[2]$ &$-1.23$&$4.1\pm0.2$&0&0&$3.9^{+0.9}_{-0.6}$&$2.1$&$27$\\
$[3]$&$-0.89\pm0.10$&$2.6\pm0.3$&0&0&$1.0^{+0.7}_{-0.8}$&$1.2$&$25$\\
$[4]$&$-1.23$&$0.15_{-0.15}^{+0.15}$&$-0.73$ &$0.72_{-0.15}^{+0.43}$&$0.45^{+1.8}_{-0.35}$&$1.3$&$26$\\
\hline
\end{tabular}
\end{center}
\end{minipage}
\end{table*}
We fit the LECS ($0.1-2$~keV), MECS ($2-10$~keV) and PDS ($15-60$~keV) data points during the first $20$~ks of the observation, by using an absorbed power-law model spectrum where we set $\rm N_{\rm H}=2.1\times10^{20}$~cm$^{-2}$, for the galactic hydrogen column density at the burst position, and account for the effect of absorption locally to the GRB site by adding $\rm N^{\rm z}_{\rm H}$ (see Table 2). The $\rm N_{\rm H}$ was modeled using the \citet{Anders1982} relative abundances. To provide for flux inter-calibration when fitting simultaneously the LECS, MECS and PDS data, the LECS/MECS and PDS/MECS normalization ratios have been varied in the range $0.7-1$ and $0.77-0.95$, respectively \citep{BeppoCookbook}. The best fit value for the spectral index is $\beta_{\rm X}=-0.82\pm0.10$ (see Table 2), consistent with the $\beta_{\rm X}=-0.94\pm0.12$ found by \citet{Maiorano2004} and obtained leaving unconstrained the range of values in which the normalizations ratio of the LECS and PDS can vary, and considering the LECS data between $0.6$~keV and $4$~keV \citep[see Fig. 3 of][]{Maiorano2004}.

Excluding the PDS data and fitting the only LECS and MECS points we get $\beta_{\rm X}=-0.89\pm0.10$, which is not significantly different from the previous result. 

The best fit values for the spectral indices relative to the next $75$~ks and to the last $70$~ks of the LECS and MECS observation are $\beta_{\rm X}=-1.2\pm0.2$ and $\beta_{\rm X}=-1.0\pm0.2$, respectively. These two values are consistent, within the errors, with the one obtained during the first $20$ ks, so they do not give any significant evidence of spectral evolution, in agreement with the results found by \citet{Maiorano2004}.
\section{Discussion}
\subsection{The prompt event in the ``forward plus reverse shock'' standard model}
\label{prompt_discussion}
The comparison of the optical, X-ray and $\gamma$-ray data presented in Fig.~\ref{fig:punti_spettrali}, shows that most of the energy is emitted in the $\gamma$-rays.  The extrapolation of the high energy time resolved spectra to optical frequencies falls at least 2 orders of magnitude below the simultaneous optical measurements, indicating the presence of an unobserved break between the optical and X-ray bands. This suggests different physical origins for the emission mechanism at low and high frequencies, confirming the idea of a reverse shock origin for GRB~990123 optical flash \citep{Sari1999b,Galama1999,Briggs1999}.

The current  standard model for $\gamma$-ray bursts and their afterglows  invokes a ``fireball'' in relativistic expansion, probably within a collimated structure (jet). The high energy burst is thought to be produced by internal shocks developing from collisions of plasma shells with different velocities. When the outflow runs into the interstellar medium (ISM), it sweeps up the surrounding gas and heats it. A reverse shock does the same to the ejecta. In these shocks, the accelerated electrons radiate via synchrotron emission. The forward external shock, interacting with the ISM, produces the multi-wavelength afterglow. 

Since the shocked region behind the reverse shock is denser and cooler than the region  behind the forward shock, it can radiate in the optical band \citep{Meszaros1997,Sari1999a}. Optical emission from the reverse shock is expected to be detected during or soon after the high energy event \citep{Sari1999b}, and was actually observed in GRB 990123 \citep{Akerlof1999}. Also the radio flare, observed at less than $3$~d after the GRBM trigger, can be interpreted as radiation from a reverse shock \citep{Kulkarni1999b} rather than afterglow from a forward shock \citep{Galama1999}. The \textit{Beppo}SAX measurements of the prompt event agree with this scenario.
\subsection{GRB~990123 afterglow}
\label{subsection:afterglow}
\subsubsection{The optical to X-ray temporal and spectral indices in the standard fireball model}\label{sec:3.2.1}
In the standard picture \citep{Sari1998}, for a spherical shock of energy $E$ propagating into a surrounding medium of density $n$, the afterglow emission $F_\nu(t)$ scales as follows: 
for $\nu$ above the cooling frequency $\nu_c$,
\begin{equation}
\label{eq:standard1}
F_\nu(t)\propto E^{(p+2)/4}t^{\alpha}\nu^{\beta}
\end{equation}
 with $\alpha=-3p/4+2/4$ and $\beta={-p/2}$ and, for $\nu<\nu_c$,
\begin{equation}
\label{eq:standard2} 
F_{\nu}(t) \propto E^{(p+3)/4}n^{1/2}t^{\alpha}\nu{^\beta}
\end{equation}
with  $\alpha=-3p/4+3/4$ and $\beta=-(p-1)/2$. Here $p$ is the power-law index of the shocked electrons.

As we said in section \ref{The X-ray afterglow}, the afterglow light curve of GRB 990123 in the $2-10$~keV energy band is well described by a power law of index $\alpha_{\rm X}=-1.46\pm0.04$. During the first day, starting from $4.1\,{\rm hr}$ after the burst, also the optical flux in the Gunn r band decreases steadily, and can be described as $F_{\rm r}=70\times(t/6\,{\rm hr})^{\alpha_{\rm opt}}\,\mu$Jy with $\alpha_{\rm opt}=-1.10\pm0.03$,  as found by \citet{Kulkarni1999a}; other authors give $\alpha_{\rm opt}=-1.12\pm0.08$ \citep{Holland2000}, $\alpha_{\rm opt}=-1.13\pm0.02$ \citep{Castro-Tirado1999} and $\alpha_{\rm opt}=-1.09\pm0.05$ \citep{Fruchter1999}, which are all consistent with the value of $-1.10\pm0.03$ that we adopt here. Two days after the burst, the optical flux declined more rapidly, with $\alpha_{\rm opt}=-1.65\pm0.06$ \citep{Kulkarni1999a}. This steepening has been ascribed either to a transition  of the fireball to a non-relativistic phase \citep{Dai1999} or to the signature of the detection of a relativistic jet \citep{Rhoads1999,Sari1999c,Meszaros1999,Huang2000a,Huang2000b,Huang2000d,Wei2000}. Since both are estimated to take place approximately $2$~d after the GRB start time, i. e. after the end of the \textit{Beppo}SAX follow-up observations, they would not be relevant for the X-ray afterglow, that can therefore be studied assuming a spherically symmetric relativistic expansion. Moreover, the faster decline at X-ray wavelengths is indicative of an evolution in a constant density environment and it is opposite to the expectations of the wind model \citep{Chevalier1999}. Thus, in the standard synchrotron fireball model, the optical and X-ray afterglow spectral and temporal indices of GRB~990123 should be compared with the closure relationships indicated in Table 3, that have been derived from equations (\ref{eq:standard1}) and (\ref{eq:standard2}).
\begin{table*}
\begin{minipage}[t]{\textwidth}
\begin{center}
\caption{Closure relationships in the standard synchrotron fireball model and their corresponding values for $\beta_{\rm opt}=-0.75\pm0.07$, $\beta_{\rm X}=-0.82\pm0.10$, $\alpha_{\rm opt}= -1.1\pm0.03$ and $\alpha_{\rm X}= -1.46\pm0.04$.} 
\begin{tabular}{c c c c}
\hline\hline
\multicolumn{4}{c}{$\nu_{\rm opt}<\nu_{\rm X}<\nu_{c}$}\\
\hline
[a] $\alpha_{\rm opt}-3/2\beta_{\rm opt}=0$&[b] $\alpha_{\rm X}-3/2\beta_{\rm X}=0$&[c] $\alpha_{\rm opt}-\alpha_{\rm X}=0$&[d] $\beta_{\rm opt}-\beta_{\rm X}=0$\\
$0.02\pm0.11$&$-0.23\pm0.15$&$0.360\pm0.050$\footnote{Not consistent with the expectations at the $7.2\sigma$ level.}&$0.07\pm0.12$\\
\hline\hline
\multicolumn{4}{c}{$\nu_{\rm opt}<\nu_{c}<\nu_{\rm X}$}\\
\hline
[a] $\alpha_{\rm opt}-3/2\beta_{\rm opt}=0$&[b] $\alpha_{\rm X}-3/2\beta_{\rm X}-1/2=0$&[c] $\alpha_{\rm opt}-\alpha_{\rm X}-1/4=0$&[d] $\beta_{\rm opt}-\beta_{\rm X}-1/2=0$\\
$0.02\pm0.11$&$-0.73\pm0.15$\footnote{Not consistent with the expectations at the $4.9\sigma$ level.}&$0.110\pm0.050$&$-0.43\pm0.12$\footnote{Not consistent with the expectations at the $3.6\sigma$ level.}\\
\hline
\end{tabular}
\end{center}
\end{minipage}
\end{table*} 

We have seen in section \ref{The X-ray afterglow} that the best fit value for the X-ray spectral index during the first $20$ ks of the NFIs observation is $\beta_{\rm X}=-0.82\pm0.10$. The power-law which best fits the optical-to-IR spectrum during the first two days has an index $-0.8\pm 0.1$ \citep{Kulkarni1999a}; this is consistent with the spectral index measured from the optical spectrum of the transient 19 hrs after the burst: $-0.69\pm 0.1$ \citep{Andersen1999}. Reconstructing the radio to X-ray afterglow spectrum on January $24.65$~UT, \citet{Galama1999} found a spectral index of $-0.75\pm0.23$ in the optical range and a spectral slope of $-0.67\pm0.02$ between the optical and the X-ray wavebands; moreover, recently \citet{Maiorano2004} have estimated a best fit optical-to-IR spectral index of $-0.60\pm0.04$. In the following discussion, we set $\beta_{\rm opt}$ equal to the mean optical-IR observed spectral index of $-0.75 \pm 0.07$, as derived by \citet{Holland2000}, and address the differences with the case $\beta_{\rm opt}=-0.60\pm0.04$ \citep{Maiorano2004} where necessary.

As shown in Table 3, if $\nu_{\rm opt}<\nu_{\rm X}<\nu_{c}$, all the closure relationships are verified within $1-2~\sigma$, except for the one relating the temporal indices of the optical and X-ray afterglow, that is not consistent with the expectations ($\sim7\sigma$ level). If we consider that GRB 990123 is one of the brightest $\gamma$-ray burst detected by the \textit{Beppo}SAX satellite and that, consequently, the temporal indices are measured with relative errors less than $5\%$, this result enables us to exclude the case of a cooling frequency above the X-ray band. 

\begin{figure*}
\centering
\includegraphics[angle=-90,width=\columnwidth]{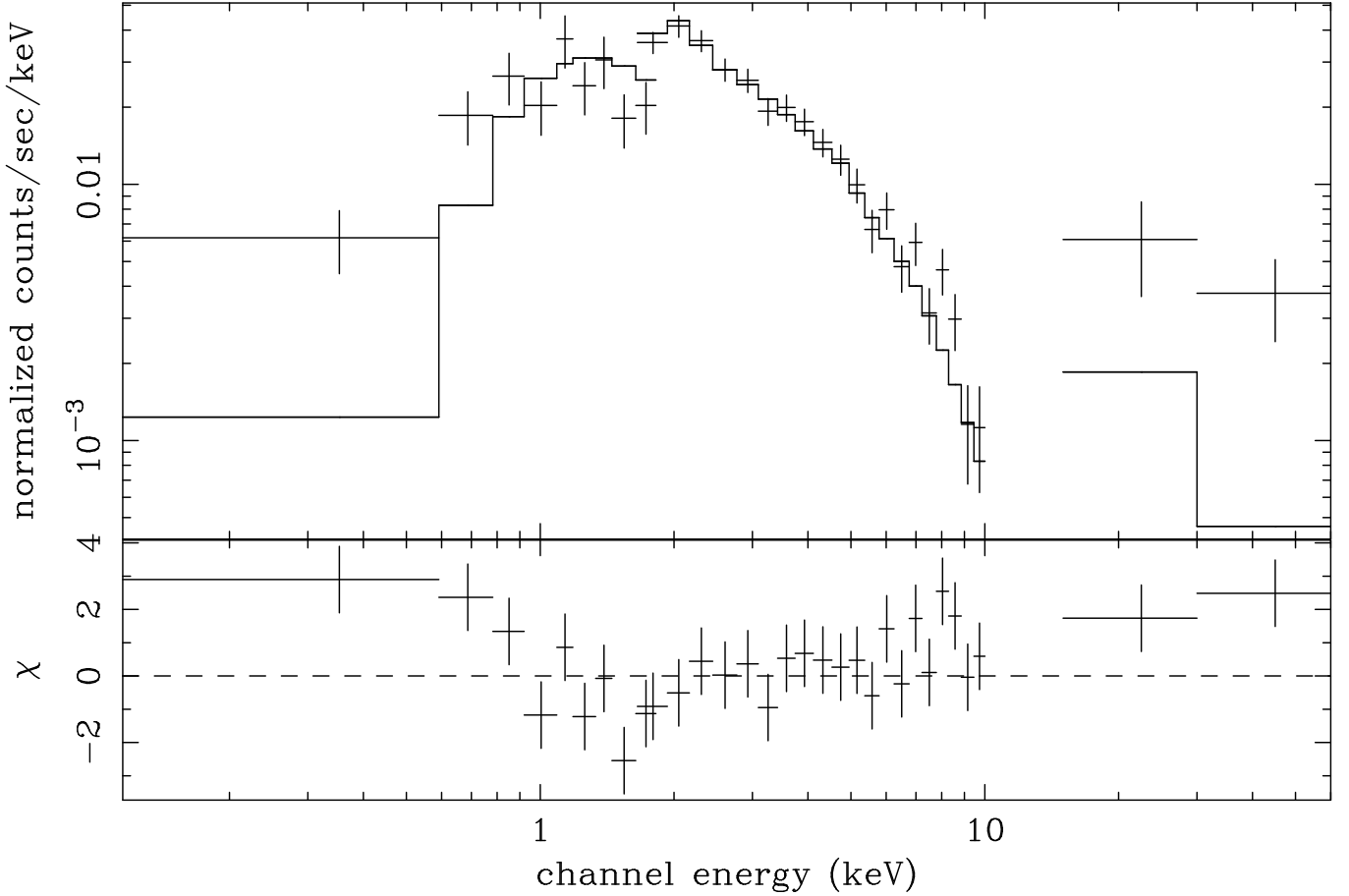}
\caption{GRB 990123 X-ray spectrum during the first $20$~ks of the NFIs observation compared with the best fit power-law model of fixed spectral index $\beta_{\rm X}=-1.23$. It is evident that the observed spectrum is too flat to be well reproduced by this model.}
\label{fig:Fnu_indice_fissato}
\end{figure*}

On the other hand, if $\nu_{\rm opt}<\nu_{c}<\nu_{\rm X}$, the temporal slopes of GRB~990123 optical and X-ray light curves are readily explained within the standard synchrotron fireball model (Table 3). However, the closure relations involving $\beta_{\rm X}$ are not consistent with the expectations, giving evidence for a too flat X-ray spectral index. This last issue can also be addressed in the following way: the observed value of $\alpha_{\rm opt}$ allows to estimate that of $p$ and, according to (\ref{eq:standard2}), gives $p=-\frac{4}{3}\alpha_{\rm opt}+1=2.47\pm0.04$; from this we get: $\alpha_{\rm X}=\alpha_{\rm opt}-1/4=-1.35\pm0.03$, $\beta_{\rm opt}=\frac{2}{3}\alpha_{\rm opt}=-0.73\pm0.02$, $\beta_{\rm X}=\frac{2}{3}\alpha_{\rm opt}-\frac{1}{2}=-1.23\pm0.02$. While the first two values are consistent with the observed ones within $2.2\sigma$ and $1\sigma$ respectively, the latter is consistent with $\beta_{\rm X}=-0.82\pm0.10$ only at the $4\sigma$ level. A spectral fit of the X-ray data points with a power-law of index fixed to $\beta_{\rm X}=-1.23$ is shown in Fig.~\ref{fig:Fnu_indice_fissato} and gives the results reported in Table 2: a similar spectral index is clearly not consistent with the data.

It is then difficult to interpret GRB~990123 afterglow within the basic synchrotron model: while for the optical afterglow both the temporal and spectral behavior are well explained by simply assuming $\nu_{c}$ between the optical and the X-ray energy band at the beginning of the \textit{Beppo}SAX follow-up observation, in the X-rays this assumption only accounts for the slope of the $2-10$~keV light curve, but not for the flatness of the $0.1-60$~keV spectrum.

Using the spectral indices found by \citet{Maiorano2004}, $\beta_{\rm X}=-0.94\pm0.12$ and $\beta_{\rm opt}= -0.60\pm0.04$, the closure relations [a], [b], [c], [d] for this case are verified within $2.8\sigma$, $3\sigma$, $2.2\sigma$ and $1.3\sigma$, respectively; thus, since the difference between the optical and the X-ray spectral index is enhanced, the observations are consistent with the relation [d], while [b] gives only marginal evidence for the existence of an X-ray excess.

Finally, we can look at the optical to X-ray normalization. As already noted by \citet{Kulkarni1999a}, at the epoch of the start of the NFIs observations the optical to X-ray spectral index appears to be rather flat, $\beta_{\rm X-opt}=-0.54\pm0.02$; this implies that the observed $2$~keV flux is somewhat higher than one would expected from the extrapolation of the optical spectrum.
\begin{figure*}
\centering
\includegraphics[angle=-90,width=\columnwidth]{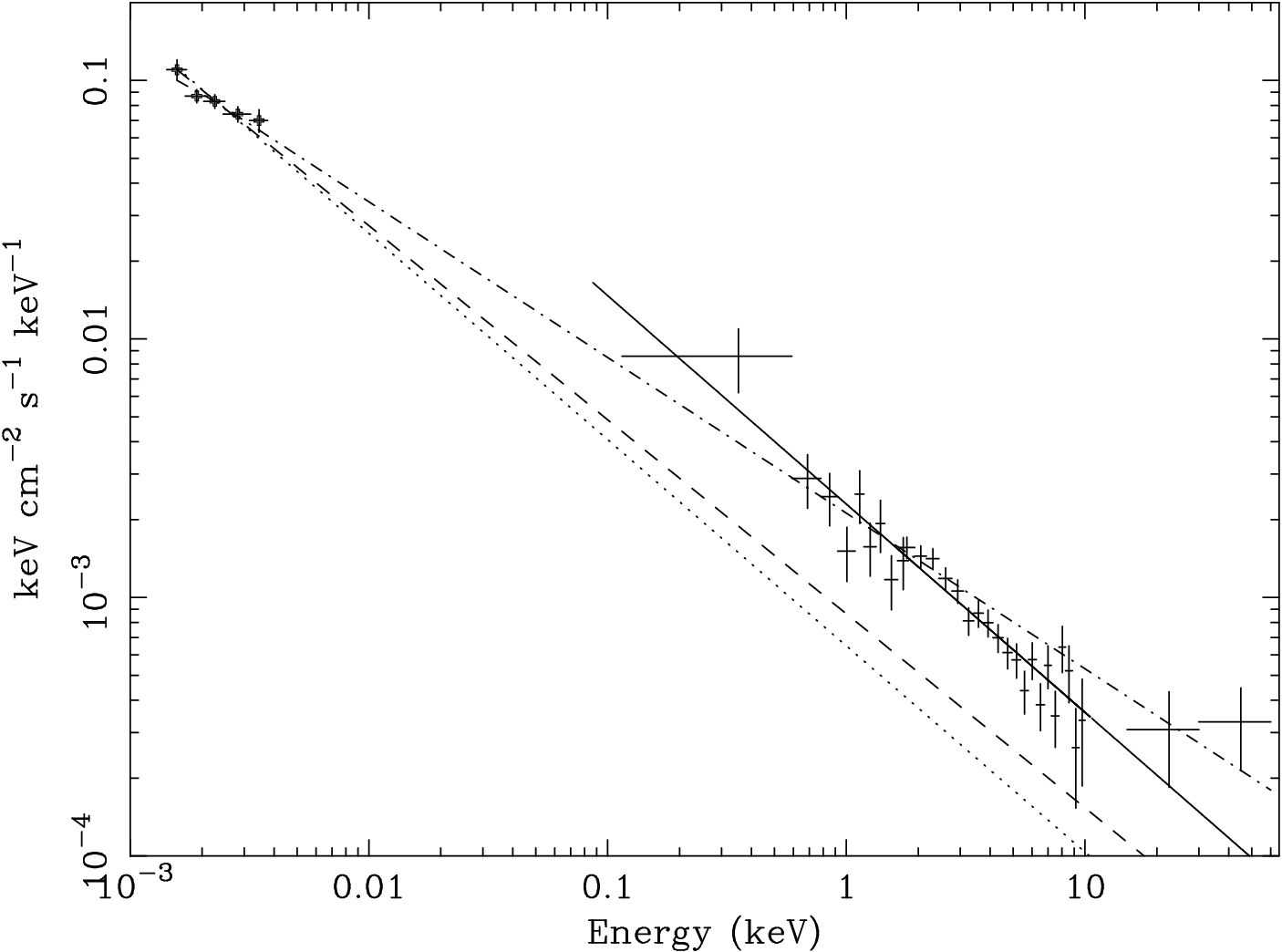}
\caption{The U, B, V, R, and I band data at $t=1999~\rm Jan.~24.497, 24.505, 24.496, 24.456, 24.487$~UT \citep{Galama1999}, extrapolated at $t\cong8.8$~hr from the burst using a temporal decay index of $\alpha_{\rm opt}=-1.1$ and extinction corrected \citep{Galama1999}, are compared with the $0.1-60$~keV spectrum corrected for galactic absorption ($\rm N_{\rm H}=2.1\times10^{20}$~cm$^{-2}$) and for absorption locally to the GRB site (see the best fit results of case $[1]$ in Table 2). The dash-dotted, dashed and dotted lines are the power-law models of fixed spectral index $\beta_{\rm opt}=-0.6$ \citep{Maiorano2004}, $\beta_{\rm opt}=-0.75$ \citep{Holland2000}, $\beta_{\rm opt}=-0.8$ \citep{Kulkarni1999a}, respectively, for the optical extrapolated data points; the solid line is the power-law model of case $[1]$ in Table 2, where $\beta_{\rm X}=-0.82$.}
\label{fig:spettro}
\end{figure*}
As shown in Fig.~\ref{fig:spettro}, assuming a mean optical spectral index of $\beta_{\rm opt}=-0.75$ \citep{Holland2000} the high energy extrapolation of the optical data falls well below the X-ray points; similarly, the low energy extrapolation of the X-ray data falls well above the optical ones, giving evidence for the presence of an X-ray excess. Assuming an optical spectral index of $\beta_{\rm opt}=-0.6$ \citep{Maiorano2004}, the optical to X-ray normalization problem is solved by including a break at a frequency $\nu\cong 1$~keV at $t\cong8.8$~hr since the trigger. On the other hand, if one assumes an optical spectral index of $\beta_{\rm opt}=-0.8$ \citep{Kulkarni1999a}, the normalization problem is moreover strengthened, suggesting that the optical and X-ray emission could be related to different components. Different cases have been found for GRB afterglows in which there is evidence for an X-ray excess of this kind, and a contribution from IC scattering have been suggested as a possible explanation. In particular, \citet{Harrison2001} analyzed the case of GRB 000926, finding that the broad-band light curves can be explained with reasonable physical parameters if the cooling is dominated by IC scattering; for this model, an excess due to IC appears above the best-fit synchrotron spectrum in the X-ray band. \citet{Castro-Tirado2003} analyzed the NIR/optical and X-ray spectra of GRB 030227, finding the two mismatching each other's extrapolation and suggesting that, in contrast to the NIR/optical band where synchrotron processes dominate, in the X-ray spectrum there could be an important contribution of IC scattering. Finally, a mismatch in the optical to X-ray spectrum was also found for the bright GRB 010222 \citep{Zand2001}. 

Considering that in the standard model the values of the temporal indices are determined by the hydrodynamical evolution of the fireball, while the shape of the broad-band spectrum depends only on the assumption of synchrotron emission, from our analysis we conclude that: (i) using the spectral indices $\beta_{\rm X}=-0.82\pm0.1$ and $\beta_{\rm opt}=-0.75\pm0.07$, the optical afterglow is well explained by synchrotron emission and a standard hydrodynamical evolution (closure relation [a]), but the X-ray one seems to require an additional flat spectrum emission component (closure relations [b] and [d]); (ii) on the other hand, using the spectral indices found by \citet{Maiorano2004}, the shape of the broad-band afterglow spectrum is consistent with a single spectral component and standard synchrotron emission (closure relation [d]), but the temporal slopes of the optical and X-ray light curves are not consistent with the observations (closure relations [a] and [b]), suggesting the possibility of a different hydrodynamical evolution. We discuss the issue (i) in the following section and (ii) in section \ref{alternativa}.
\subsubsection{A possible IC-scattering contribution: the synchrotron+IC model}
In order to interpret the combined spectral and temporal properties of GRB~990123 afterglow, we consider the possibility of relating the X-ray emission to an important contribution of IC scattering. 

The X-ray spectral index, $\beta_{\rm X}=-0.82\pm0.10$, similar to the optical one, $\beta_{\rm opt}=-0.75\pm0.07$, and the difference between the optical to X-ray temporal slopes $\alpha_{\rm opt}-\alpha_{\rm X}\cong0.36$, suggest the presence of an additional hard component. This naturally leads to a model in which the X-ray emission is dominated by IC-scattering of lower energy photons, while the optical one is synchrotron dominated; furthermore, an IC contribution could also explain the unusual high-energy ($15-60$~keV) emission observed by the PDS during the first $20$~ks of the afterglow observations. 
 
In a synchrotron+IC model, the afterglow spectrum at some time $t_{0}$ would be the sum of two power-laws:
\begin{equation}
\label{eq:s+Ic}
F_{\nu}(t_{0})=\exp\left[-\sigma(\nu){\rm N_{H}}-\sigma((1+z)\nu){\rm N_{H}^{z}}\right]\times\left[F^{\rm syn}_{0}\left(\frac{\nu}{2\rm keV}\right)^{\beta_{X}}+F^{\rm IC}_{0}\left(\frac{\nu}{2\rm keV}\right)^{\beta^{\rm IC}_{\nu}}\right]
\end{equation} with $\beta_{\rm X}$ the spectral index for the synchrotron component; $\beta_{\nu}^{\rm \rm IC}=\frac{1}{3}$ if $\nu<\nu_{\rm m}^{\rm \rm IC}(t_{0})$, where  $\nu_{\rm m}^{\rm \rm IC}$ is the peak frequency of the IC spectrum, otherwise, $\beta_{\nu}^{\rm IC}=\frac{-(p-1)}{2}$.

If the fireball expands in a medium of constant number density, if the expansion is spherically symmetric and if we are in the slow-cooling IC dominated phase \citep{Sari2001}, the contribution of IC scattering at a given frequency will evolve with time according to the following relations:
\begin{eqnarray}
F^{\rm IC}_{\nu}\propto t^{9/4}; && \nu<\nu^{\rm IC}_{a}
\end{eqnarray}
\begin{eqnarray}
F^{\rm IC}_{\nu}\propto t; && \nu^{\rm IC}_{a}<\nu<\nu^{\rm IC}_{m}
\end{eqnarray}
\begin{eqnarray}
F^{\rm IC}_{\nu}\propto t^{-(9p-11)/8}; && \nu^{\rm IC}_{m}<\nu<\nu^{\rm IC}_{c}
\end{eqnarray}
\begin{eqnarray}
F^{\rm IC}_{\nu}\propto t^{-(9p-10)/8+(p-2)/(4-p)}; && \nu>\nu^{\rm IC}_{c}
\end{eqnarray}
assuming a power-law approximation for the IC spectrum \citep{Sari2001}.
 
Observing the above spectral and temporal relations, and considering the analysis we have done in section \ref{sec:3.2.1}, it is clear that, if we want the synchrotron emission to be dominant in the optical band and the IC contribution to explain all the X-ray emission, we should have: (i) a cooling frequency of the synchrotron component above the optical band, since this condition explains well both the spectral and the temporal behavior of the optical afterglow (section \ref{sec:3.2.1}); (ii) a peak frequency of the IC component below the X-ray energy band, in order to account for the decreasing $2-10$~keV light curve (if $\nu^{\rm IC}_{m}>10$~keV, the $2-10$~keV flux would increase with time, if dominated by IC emission). 

We can also notice that in the slow-cooling IC-dominated phase, the IC contribution affects the temporal slope of the synchrotron component at frequencies above the cooling one, changing the temporal decay index from $-\frac{3(p-1)}{4}-\frac{1}{4}$ to $\frac{-3p^{2}+12p-4}{4(p-4)}$ \citep[see ][equation (
A.20) in the electronic version of this paper]{Sari2001}. This means that the difference between the temporal decay index of the synchrotron flux at frequencies below and above the cooling one becomes $\frac{1}{4}-\frac{p-2}{8-2p}$ instead of $\frac{1}{4}$: the presence of strong IC cooling makes the difference smaller than the case of pure synchrotron emission. Therefore, the case in which the IC contribution dominates the $2-10$~keV emission seems to be favored with respect, for example, to one in which the IC component becomes dominant only in the high energy tail of the X-ray spectrum (say, for $\nu\geq10$~keV), with the $2-10$~keV emission still being dominated by synchrotron emission. In fact, if this was the case, we should have $\alpha_{\rm opt}=-\frac{3(p-1)}{4}$, $\alpha_{\rm X}=\frac{-3p^{2}+12p-4}{4(p-4)}$ and $\alpha_{\rm opt}^{2}-\alpha_{\rm X}\alpha_{\rm opt}+\frac{3}{2}\alpha_{\rm opt}-\frac{9}{4}\alpha_{\rm X}-\frac{15}{16}=0$. For GRB 990123, the latter has a value of $0.301\pm0.085$, consistent with the expected one only within $3.5\sigma$. 

For this reason, we analyze the hypothesis of the IC emission dominating the whole $2-60$~keV spectrum. For $\nu_{\rm opt}<\nu_{c}<\nu_{\rm X}$ and $\nu_{\rm X}>\nu^{\rm IC}_{m}$, with the synchrotron component dominating the optical emission and the IC component the X-ray one, the closure relations should be those indicated in Table 4, that for GRB 990123 are all verified within $1\sigma$.
\begin{table*}
\begin{minipage}[t]{\textwidth}
\begin{center}
\caption{Closure relationships between the temporal and spectral indices in the synchrotron+IC model and their corresponding values for $\beta_{\rm opt}=-0.75\pm0.07$, $\beta_{\rm X}=-0.82\pm0.10$, $\alpha_{\rm opt}= -1.1\pm0.03$ and $\alpha_{\rm X}= -1.46\pm0.04$.} 
\begin{tabular}{c c c c}
\hline\hline
\multicolumn{4}{c}{$\nu_{\rm opt}<\nu_{\rm c}<\nu_{\rm X}$ and $\nu_{\rm X}>\nu^{\rm IC}_{m}$}\\\hline
$\alpha_{\rm opt}-3/2\beta_{\rm opt}=0$&$\alpha_{\rm X}-9/4\beta_{\rm X}-1/4=0$&$\alpha_{\rm opt}-2/3\alpha_{\rm X}+1/6=0$&$\beta_{\rm opt}-\beta_{\rm X}=0$\\
$0.02\pm0.11$&$0.13\pm0.23$&$0.040\pm0.040$&$0.07\pm0.12$\\
\hline
\end{tabular}
\end{center}
\end{minipage}
\end{table*}

Using the relations $\alpha_{\rm opt}=-1.10\pm0.03=-3(p-1)/4$ and $\alpha_{\rm X}=-1.46\pm0.04=-(9p-11)/8$, we get for the index of the electron energy distribution $p=2.47\pm0.02$ and $2.52\pm0.03$, respectively. Using the spectral indices $\beta_{\rm opt}=-0.75\pm0.07=-(p-1)/2$ and $\beta_{\rm X}=-0.82\pm0.10=-(p-1)/2$ we obtain: $p=2.50\pm0.14$ and $p=2.64\pm0.20$, respectively. These values are all consistent with each other within $\sim 1\sigma$, as expected from the closure relations being verified at the $1\sigma$ level. We then choose the one with the smaller error, that is $p=2.47\pm0.02$, derived from the optical temporal index $\alpha_{\rm opt}$. In this way, we estimate the expected spectral indices between $0.1-60$~keV for both the synchrotron component and the IC one; these are: $\beta_{\rm X}=2/3\alpha_{\rm opt}-1/2=-1.23\pm0.02$, and $\beta^{\rm IC}_{\rm X}=2/3\alpha_{\rm opt}=-0.73\pm0.02$, respectively.\begin{figure*}
\centering
\includegraphics[angle=-90,width=\columnwidth]{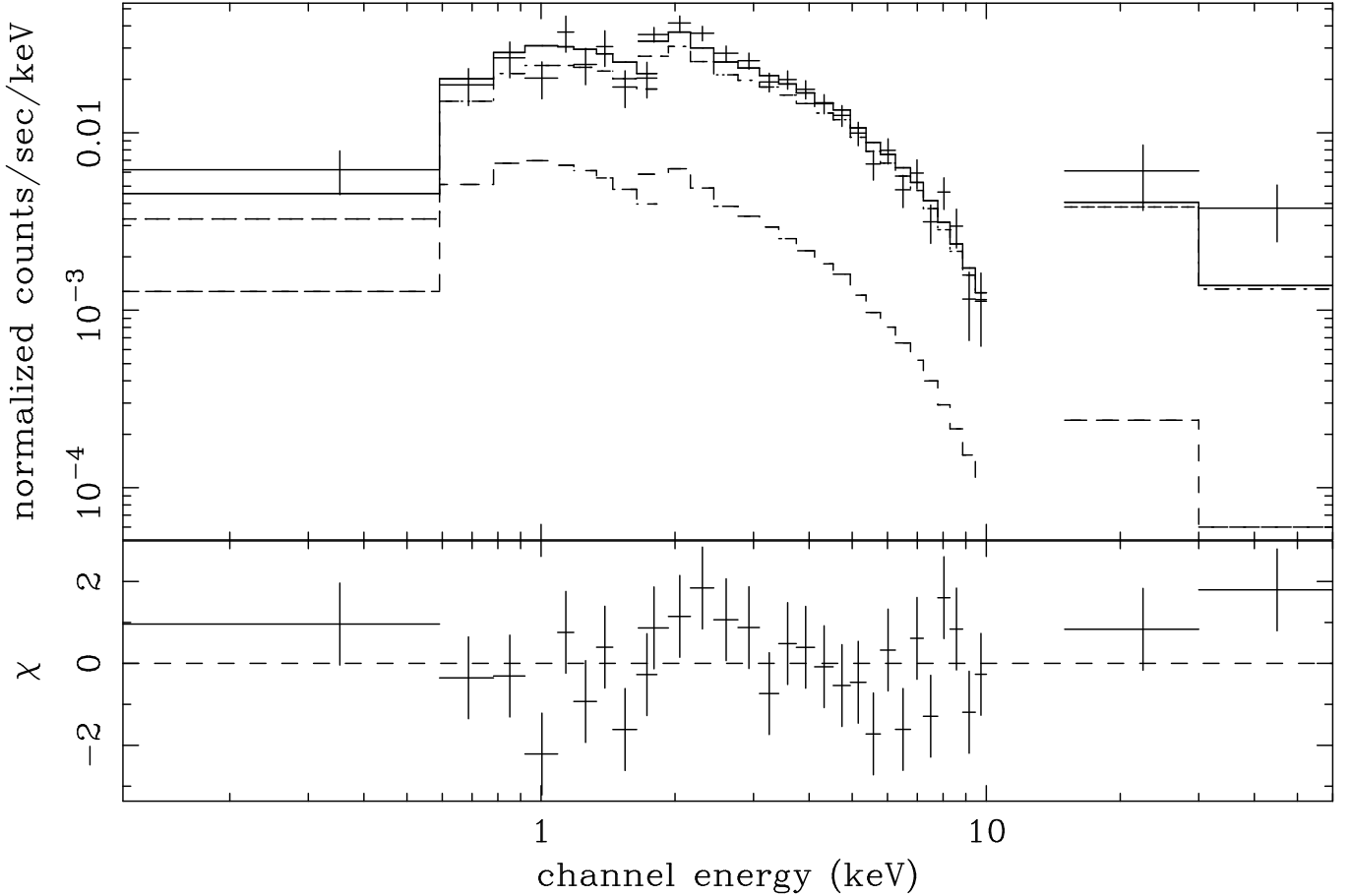}
\caption{GRB 990123 X-ray spectrum during the first $20$~ks of the NFIs observation. The model spectrum is the sum of two power-law components (see equation (\ref{eq:s+Ic})): one represents the synchrotron contribution and has a fixed spectral index $\beta_{\rm X}=-1.23$, the other accounts for IC scattering and has a fixed spectral index $\beta^{\rm IC}_{\rm X}=-0.73$.}
\label{fig:Fnu_syn+IC}
\end{figure*}
\begin{figure*}
\centering
\includegraphics[angle=-90,width=\columnwidth]{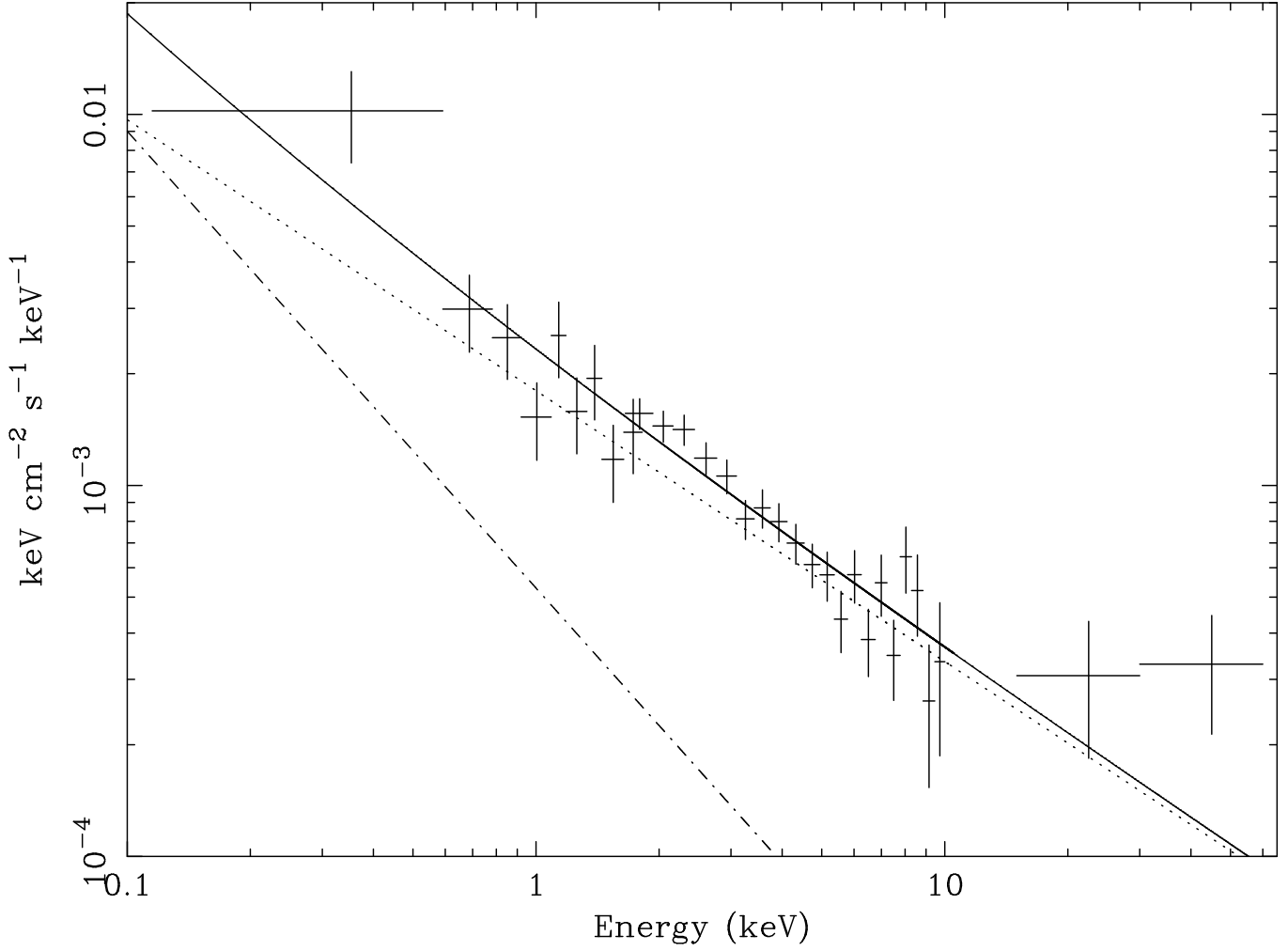}
\caption{GRB 990123 $F_{\nu}$ X-ray spectrum: the data and the model are those of Fig.~\ref{fig:Fnu_syn+IC} corrected for LECS/MECS and PDS/MECS flux inter-calibration normalization ratios and for absorption.}
\label{fig:nuFnu_syn+IC}
\end{figure*}

In Fig.~\ref{fig:Fnu_syn+IC} and Fig.~\ref{fig:nuFnu_syn+IC} we fit the $0.1-60$~keV data during the first $20$~ks of the NFIs observation with a model spectrum of the form (\ref{eq:s+Ic}), where we set $\beta_{\rm X}=-1.23$ and $\beta_{\rm X}^{\rm IC}=-0.73$. The best fit values (see Table 2) for the synchrotron and IC flux at $2$~keV are $F_{0}^{\rm syn}=0.15^{+0.15}_{-0.15}~\mu$Jy and $F_{0}^{\rm IC}=0.72^{+0.43}_{-0.15}~\mu$Jy ($\chi^{2}=35.5/26$), respectively, consistent with the hypothesis of an IC component dominating on the synchrotron one in the X-ray band: $F_{0}^{\rm syn}/F_{0}^{\rm IC}\leq0.15/0.72\cong0.21$, taking into account that for the synchrotron flux at $2$~keV we get only an upper limit. 
 
Our following step is to relate $\nu_{m}^{\rm IC}$, $\nu_{c}$, $F_{2\rm keV}^{\rm syn}$ and $F_{m}^{\rm IC}$ at $t=0.37$~d to the fundamental parameters of the fireball model: the fraction of the shock energy density that goes into the electrons and magnetic energy density, $\epsilon_{e}$ and $\epsilon_{B}$, respectively; the energy of the fireball  $E_{52}$ in units of $10^{52}$~ergs; the ambient medium number density $n_{1}$, in units of particles$/$cm$^{3}$. 

Using the formulas given by \citet{Granot1999a}, \citet{Sari1998}, and by \citet{Sari2001} in the slow-cooling IC dominated regime (see the Appendix in the electronic version of this paper for details), for the synchrotron cooling frequency we can write:
\begin{equation}
\nu_{c}({\rm Hz})=A_{\nu_{c}}\left(\epsilon_{B}E_{52}\right)^{a}n_{1}^{b}\epsilon_{e}^{c}
\label{numero1}
\end{equation}
where:
\begin{equation}
A_{\nu_{c}}({\rm Hz})=(2.7\times10^{12})(1+z)^{\frac{p}{2(p-4)}}(170)^{\frac{p-2}{p-4}}t_{d}^{\frac{8-3p}{2(p-4)}}
\end{equation}
with $t_{d}$ the time since the burst measured in the observer's frame in units of days, and:\begin{equation}
a=\frac{p}{2(p-4)}
\end{equation}
\begin{equation}
b=\frac{2}{p-4}
\end{equation}
\begin{equation}
c=2\left(\frac{p-1}{p-4}\right)
\end{equation}
For the peak frequency of the IC component:
\begin{equation}
\nu^{\rm IC}_{m}({\rm Hz})=A_{\nu^{\rm IC}_{m}}\epsilon_{B}^{1/2}\epsilon_{e}^{4}E_{52}^{3/4}n_{1}^{-1/4}\label{numero2}\end{equation}
where:
\begin{equation}
A_{\nu_{m}^{\rm IC}}({\rm Hz})=2(2.9\times10^{15})(1.116\times10^{4})^{2}(1+z)^{1/2}\left(\frac{p-2}{p-1}\right)^{4}t_{d}^{-9/4}
\label{eq:ampiezza_nu_IC_m}
\end{equation}
For the peak flux of the IC component:
\begin{equation}
F_{m}^{\rm IC}(\mu{\rm Jy})=A_{F_{m}^{\rm IC}}\left(E_{52}n_{1}\right)^{5/4}\epsilon_{B}^{1/2}
\label{eq:flusso_max_IC}
\end{equation}
where:
\begin{equation}
A_{F_{m}^{\rm IC}}(\mu{\rm Jy})=0.39(2\times10^{-7})(1.7\times10^{4})(1+z)d_{28}^{-2}t_{d}^{1/4}\end{equation}
with $d_{28}$ the luminosity distance of the source in units of $10^{28}$~cm.
For the flux of the synchrotron component at a frequency $\nu>\nu_{c}$:
\begin{equation}
F_{\nu}(\mu{\rm Jy})=A_{F_{\nu}}\epsilon_{B}^{d}n_{1}^{e}E_{52}^{f}\epsilon_{e}^{g}
\end{equation}
where:
\begin{equation}
A_{F_{\nu}}(\mu{\rm Jy})=(1.7\times10^{4})(2.9\times10^{15})^{\frac{p-1}{2}}(2.7\times10^{12})^{1/2}\nu({\rm Hz})^{-p/2}d_{28}^{-2}\left(\frac{p-2}{p-1}\right)^{p-1}(170)^{\frac{p-2}{2(p-4)}}(1+z)^{\frac{p^{2}-12}{4(p-4)}}t_{d}^{\frac{-3p^{2}+12p-4}{4(p-4)}}
\end{equation}
\begin{equation}
d=\frac{p^{2}-2p-4}{4(p-4)}
\end{equation}
\begin{equation}
e=\frac{p-2}{2(p-4)}
\end{equation}
\begin{equation}
f=\frac{p^{2}-12}{4(p-4)}
\end{equation}
\begin{equation}
g=\frac{p^{2}-4p+3}{p-4}
\end{equation}
Setting $p=2.47$, $z=1.6004$, $d_{28}=3.7$ \citep{Kulkarni1999a}, $t_{d}=0.37$~d, we have verified, by inverting the above relations (see the calculations in the Appendix), if reasonable values of $\epsilon_{B}$, $\epsilon_{e}$, $E_{52}$, and $n_{1}$ can be found when $\nu_{c}\geq2$~eV and $\nu_{m}^{\rm IC}(0.37$~d$)\le2.0$~keV.

In order to minimize the energy requirement (see equation (
A.61)), it is convenient to set (i) $\nu_{m}^{\rm IC}(0.37$~d$)\cong2$~keV and (ii) $\nu_{c}\cong2$~eV, that is to say to set these frequencies at the higher and lower edge of the range in which they can vary to be consistent with the X-ray and optical data, respectively. These conditions are immediately evident taking into account that $F_{2 \rm keV}^{\rm IC}=F_{m}^{\rm IC}(2 \rm keV/\nu_{m}^{\rm IC})^{-\frac{p-1}{2}}$, so $F_{m}^{\rm IC}$ can be expressed as a function of $\nu_{m}^{\rm IC}$ and $F_{2 \rm keV}^{\rm IC}$, the value of which is fixed from the observations; substituting into equation (
A.61), yields $E_{52}\propto(\nu_{c})^{a_{13}}(\nu_{m}^{\rm IC})^{a_{14}-a_{15}\frac{p-1}{2}}$, and for $p=2.47$: $E_{52}\propto\nu_{c}(\nu_{m}^{\rm IC})^{-0.36}$. Thus, minimizing $\nu_{c}$ and maximizing $\nu_{m}^{\rm IC}$ gives the smallest value for $E_{52}$.  

With the assumption (i) we have $F^{\rm IC}_{m}=F^{\rm IC}_{2 \rm keV}$; moreover, because of (ii), the synchrotron contribution at $2$~keV extrapolated from the Gunn r data should be $F^{\rm syn}_{2~\rm keV}\cong70(8.8~\rm hr/6~\rm hr)^{-1.46}(2~\rm keV/2~\rm eV)^{-1.23}\cong10^{-2}~\mu$Jy, which is consistent with the X-ray data best fit value of $F_{0}^{\rm syn}=0.15^{+0.15}_{-0.15}~\mu$Jy within the errors. 

Fitting the $0.1-60$ keV spectrum with model (\ref{eq:s+Ic}) where $\beta_{\rm X}=-1.23$, $\beta_{\rm X}^{\rm IC}=-0.73$ and $F_{0}^{\rm syn}=10^{-2}~\mu$Jy, yields $F^{\rm IC}_{0}=F^{\rm IC}_{2~\rm keV}=F^{\rm IC}_{m}=0.83\pm0.04~\mu$Jy. We thus set $\nu_{c}=2$~eV, $F_{0}^{\rm syn}=10^{-2}~\mu$Jy, $\nu_{m}^{\rm IC}=2~\rm keV$, $F^{\rm IC}_{m}=0.83~\mu$Jy, that give:
\begin{eqnarray}
\label{eq:epsilone_epsilonB}
\epsilon_{B}\cong10^{-10}&\epsilon_{e}\cong10^{-1}	
\end{eqnarray}
\begin{eqnarray}
E_{52}\cong7\times10^{4}&n_{1}\cong10^{2}
\label{eq:E_n}
\end{eqnarray}
With those values for the intrinsic parameters of the fireball, the temporal and spectral properties of the optical to X-ray afterglow are explained within the fireball model and the optical to X-ray normalization problem is solved. 

A similar calculation can be done setting $F_{0}^{\rm syn}=0.15~\mu$Jy and $F_{m}^{\rm IC}=0.72~\mu$Jy, that are the values we get from the X-ray data. This choice yields a higher value for $\epsilon_{B}$ and a lower one for $n_{1}$:
\begin{eqnarray}
\epsilon_{B}\cong6\times10^{-8}&\epsilon_{e}\cong5\times10^{-2}	
\end{eqnarray}
\begin{eqnarray}
E_{52}\cong6\times10^{4}&n_{1}\cong8
\end{eqnarray}
but has the problem that it predicts an optical extrapolated flux about $15$ times above the observed one ($0.15~\mu$Jy$/10^{-2}~\mu$Jy~$\cong15$). The hypothesis of a ``gray absorber'' could explain an optical flux lower than expected from the X-ray data, without affecting the optical spectral slope, and alternative scenarios to explain simultaneously the optical and X-ray data of some GRBs by invoking a more gray extinction have recently been proposed by \citet{Stratta2004}, for those GRBs for which some evidence of absorption locally to the GRB site was found. In our case, we do not have a strong evidence for local absorption (see Table 2), but we can neither exclude this hypothesis at all.

We can also notice that with the values (\ref{eq:epsilone_epsilonB}), being $\epsilon_{e}/\epsilon_{B}\geq10$, IC scattering dominates the total cooling over the whole relativistic stage of the afterglow evolution, as underlined by \citet{Sari2001}; thus, the transition to the slow-cooling synchrotron dominated phase, in which the IC cooling rate is weaker than the synchrotron one and therefore has no effects on the synchrotron spectrum, should not be observable. This implies that for the $0.1-60$ keV spectra relative to next $75$ ks and last $70$ ks, we do not expect changes in the spectral slope, in agreement with the results presented in section \ref{The X-ray afterglow}. However, we of course expect the flux level of the following spectra becoming dimmer of a factor determined by the slope of the X-ray light curve. Since the $15-60$ keV upper limits are in fact consistent with a decay slope of $\alpha=-1.46\pm0.04$ \citep[see Fig. 1 in][]{Maiorano2004}, we thus conclude that the lack of a significant detection by the PDS in the following $75$ ks and last $70$ ks spectra, has to be related to poorer statistics rather than to a strong suppression of the IC contribution.

Since we are before $2$~d from the burst, we have used the isotropic dynamical equations and $E_{52}$ is the fireball isotropic equivalent energy at the beginning of the afterglow phase. Considering that the observed isotropic equivalent $\gamma$-ray energy, $E_{\rm iso}(\gamma)=4\pi F_{\gamma}d^{2}_{L}(1+z)^{-1}$, is $1.2\times10^{54}$~ergs, the efficiency $\eta_{\gamma}$ of the fireball in converting the energy in the ejecta $E^{0}_{\rm iso}$ into $\gamma$-rays has to be $\frac{E_{\rm iso}(\gamma)}{E^{0}_{\rm iso}}=\eta_{\gamma}\cong0.2\%$, a reasonable value according to \citet{Panaitescu1999} and \citet{Beloborodov2000}.

To estimate the energy of the jet $E_{\rm j}$ at the beginning of the afterglow phase, we need the value of the jet opening angle $\theta_{\rm j}$. According to \citet{Frail2001}:
\begin{equation}
\theta_{\rm j}=0.057\left(\frac{t_{\rm j}}{1 \rm d}\right)^{3/8}\left(\frac{1+z}{2}\right)^{-3/8}\left[\frac{0.2\times E^{0}_{\rm iso}}{10^{53}\rm ergs}\right]^{-1/8}\left(\frac{n}{0.1 \rm cm^{-3}}\right)^{1/8}\end{equation}
Since $t_{\rm j}\cong 2$~d, we obtain $\theta_{\rm j}=0.064\cong3.7^{\circ}$; this implies $E_{\rm j}\cong10^{54}$~ergs for the energy in the jet.

This estimate on the jet opening angle is based on the simple relation $\gamma(t_{\rm j})=\theta^{-1}_{\rm j}$, that itself cannot distinguish between the structure of GRBs jets \citep{Meszaros1998,Dai2001,Rossi2002,Zhang2002}. Conventionally, the late time temporal index $\alpha_{2}$ for the optical afterglow of jetted GRBs, is believed to be the same as $p$. However, there are some caveats on this assumption \citep[e.g. ][]{Wu2004}, most importantly the fact that the ambiguity of the understanding on the sideways expansion of the jet leads to a great uncertainty on the value of $\alpha_{2}$. More specifically, as explained by \citet{Meszaros1999} for the case of GRB 990123, if one assumes that the steepening in GRB 990123 optical afterglow from $t^{-1.1}$ to $t^{-1.65}$ is caused by sideways expansion of the decelerating jet \citep{Rhoads1997}, one expects a large steepening from $t^{-1.1}$ to $t^{-p}$ (where $p=2.47$ in our case), more than one power of $t$. However, the edge of the jet begins to be seen when $\gamma$ drops below the inverse jet opening angle $1/\theta_{\rm j}$. This occurs well before the sideways expansion starts \citep{Panaitescu1998} and the latter is unimportant until the expansion is almost non-relativistic: GRB 990123, as well as other GRBs like GRB 010222, 020813, 021004, 000911 \citep{Wu2004}, are good candidates for non-lateral expansion jets, from which we expect a steepening by $t^{3/4}$ in the late time optical light curve. The observed difference of $\alpha_{1}-\alpha_{2}=0.55\pm0.07$, matches the expectations at the $2.8\sigma$ level and is thus in agreement with the expected change in the decay slope from seeing the edge of the jet within a model of typical electron index of $p=2.5$ \citep{Meszaros1999}. These arguments are sufficient to justify the hypothesis of a jet structure within a synchrotron+IC model with $p=2.47$, but there are of course several other plausible causes for steepening \citep[e. g. ][]{Dai1999}.
       
We can now compare our results with some previous analysis, bearing that those were not constrained by the $0.1-60$~keV afterglow spectrum, which is analyzed in this work.
For what concerns the total energy in the fireball, a value of the order of $10^{4}$ for $E_{52}$ has recently been proposed by \citet{Panaitescu2004} analyzing GRB~990123 early optical emission in the frame of the standard ``reverse-forward shock'' model. 

In \citet{Panaitescu2001}, fitting the light curves in the radio, optical and X-ray energy band, it was proposed $p\cong2.2\div2.4$, $\epsilon_{e}\cong9\times10^{-2}\div2\times10^{-1}$, $\epsilon_{B}\cong10^{-4}\div3\times10^{-3}$, $n_{1}\cong3\times10^{-4}\div2\times10^{-3}$, $\theta_{\rm j}\cong1^{\circ}\div3^{\circ}$ and $ E_{\rm j}\cong10^{50}\div5\times10^{50}$~ergs \citep[values taken from Fig. 2 in ][]{Panaitescu2001}; in particular, for $E_{\rm j}=5\times10^{50}$~ergs and $\theta_{\rm j}=3^{\circ}$, the isotropic equivalent energy is of about $4\times10^{53}$~ergs, $30$ times less than the lower limit on the isotropic energy ($E\geq10^{55}$~ergs) found by \citet{Panaitescu2004} and five times less than the observed isotropic energy emitted in $\gamma$-rays. A value of $4\times10^{53}$~ergs requires the existence of a radiative evolution phase before the adiabatic one, causing the emission of more of the initial explosion energy and leaving less for the adiabatic phase. The same considerations can also be done for the model proposed by \citet{Wang2000}, who set $p=2.44$, $\epsilon_{e}\cong0.57$, $\epsilon_{B}\cong3.1\times10^{-3}$, $n_{1}\cong0.01$, $E_{52}\cong5$ and considered only synchrotron emission: in this case $E_{52}$ is about $30$ times less than $E_{\rm iso}(\gamma)$.

Finally, about the power-law index of the electron energy distribution, values of $p=2.44$ and $p=2.3$ were found by \citet{Wang2000} and \citet{Panaitescu2001} respectively, similar with the one of $p=2.47$ we suggest in our analysis. In this work, however, to fit the observed flat $0.1-60$~keV spectrum and give an overall interpretation of the multi-wavelength afterglow, we are adding a dominant contribution of IC up-scattering to the synchrotron emission process. 

\subsubsection{The radio observations: problems of the synchrotron+IC model}
GRB~990123 has been largely discussed within the standard fireball theory and a reverse shock origin has been assigned to both the optical flash and the radio flare \citep{Sari1999a,Galama1999,Briggs1999,Kulkarni1999b}. The light curves in the radio, optical and X-ray energy bands have been well reproduced \citep{Panaitescu2001,Kulkarni1999b} and different sets of values for the intrinsic parameters $\epsilon_{e}$, $\epsilon_{B}$  $E_{52}$ and $n_{1}$ have been proposed \citep{Panaitescu2001,Panaitescu2004,Wang2000}. 

In this paper we are adding another piece of information to the puzzle: the spectrum of the optical to X-ray afterglow extended to the high energy tail of the $0.1-60$~keV energy band. As we have underlined in the previous sections, GRB~990123 appears anomalous and difficult to interpret within a standard synchrotron model when combining this spectral information with the temporal properties of the optical and X-ray light curves. This is what has brought us to study the hypothesis of a synchrotron+IC model with values for the intrinsic parameters (\ref{eq:epsilone_epsilonB}) and (\ref{eq:E_n}).

We want now to verify if such a model is consistent with the $8.46$~GHz data. The radio afterglow was observed from $t>1$~d after the burst \citep[here we are not considering the data relative to the radio flare, to which we assign a reverse shock origin, ][]{Kulkarni1999b}. To estimate the value of the $8.46$~GHz flux at about $1$~d from the burst, we need to calculate the value of the synchrotron self-absorption frequency.
According to \citet{Granot1999b}, the self-absorption frequency $\nu_{a}$ is:
\begin{equation}
\nu_{a}({\rm Hz})=0.247\times4.24\times10^{9}(1+z)^{-1}\left(\frac{p+2}{3p+2}\right)^{3/5}\frac{(p-1)^{8/5}}{p-2}\epsilon_{e}^{-1}\epsilon_{B}^{1/5}E_{52}^{1/5}n_{1}^{3/5}\end{equation}
Using the values (\ref{eq:epsilone_epsilonB}) and (\ref{eq:E_n}) we get $\nu_{a}\cong15$~GHz. 

Further, substituting those values in equations (
A.1) and (
A.2) of the Appendix, we get $\nu_{m}(1~{\rm d})\cong13$~GHz~$\cong\nu_{a}$ and $f_{m}^{\rm syn}\cong23$~mJy. Thus we have:
\begin{equation}
F_{8.46\rm GHz}(1~{\rm d})\cong23~{\rm mJy}\left(\frac{8.46~\rm GHZ}{15~\rm GHz}\right)^{2}\cong7~\rm mJy
\end{equation}
that is about a factor of $30$ above the observations, that are lower than $260$~$\mu$Jy \citep{Kulkarni1999b}.

Another way to explain the problem is the following: over the first two days, the most abundant photons are between the radio and the optical regimes; if these photons are up-scattered by the Compton process, then the required Thompson optical depth is $\tau=f_{\rm X}/f_m\cong0.83~\mu$Jy$/F^{\rm syn}_{m}$. The observed radio/optical fluxes in the case of GRB~990123 are lower than $\sim260~\mu$Jy \citep{Kulkarni1999b}, that imply $\tau\geq0.83/260\cong 3\times 10^{-3}$. On the other hand, the optical depth has the following dependence on the parameters of the fireball model:
\begin{equation} 
\tau=\sigma_T n R\cong 4 \times 10^{-7} (E/10^{54}{\rm erg})^{1/4} (n/1{\rm cm}^{-3})^{3/4}(t/6 {\rm hr})^{1/4}
\end{equation}
which has a value of $\sim10^{-4}$, for $E_{52}=7\times10^{4}$, $n_{1}=100$ and $t\cong8.8$~hr. Thus, to reconcile this value of $\tau$ with the observations, one should have $F_{m}^{\rm syn}$ of the order of $3\times10^{-3}/10^{-4}\times260~\mu$Jy, that is to say 30 times higher than the observed one, that is the ultimate result of our synchrotron+IC model. 

\subsubsection{An alternative solution for the puzzle}
\label{alternativa}
 In the previous sections we noticed that the observed temporal slopes of the optical and X-ray light curves of GRB~990123 afterglow cannot be reconciled with the observed value of the X-ray spectral index during the first $20$~ks of the NFIs observation, at least in the frame of the basic synchrotron standard model. Thus we tried to explain the X-ray spectrum by invoking a significant contribution of another emission mechanism in addition to the synchrotron one, that is IC scattering. However, we have some problems when comparing our results with the $8.46$~GHz upper limits. 

A completely different approach could be to leave unchanged the emission mechanism, but change the hydrodynamics. Assuming that synchrotron radiation from electrons with a power law energy distribution is the only efficient mechanism, the shape of the afterglow spectrum is independent of the details of the hydrodynamics, thus the relation between $p$ and $\beta$ is fixed. From the optical to X-ray spectral index, $\beta_{\rm X-opt}=-0.54\pm0.02$ \citep{Kulkarni1999a}, we can infer $p=2.08\pm0.04$ if $\nu_{c}\cong1$~keV; this value of $p$ predicts $\beta_{\rm opt}=-0.54\pm0.02$ and $\beta_{\rm X}=-1.04\pm0.02$, that agree at the $\sim 1\sigma$ level with the spectral indices found by \citet{Maiorano2004}. Assuming $\beta_{\rm opt}=-0.75\pm0.07$ and $\beta_{\rm X}=-0.82\pm0.10$, the agreement is at the $2\sigma$ level. However, in this case, the closure relation $\beta_{\rm opt}-\beta_{\rm X}-1/2=0$ is only verified at the $3.6\sigma$ level. 

Further, in order to account for the temporal decay (a value of $p=2.08$ predicts an X-ray temporal index of $-1.06\pm0.03$ and an optical temporal index of $-0.81\pm0.03$, that are not consistent with the observed ones), we need to change the
hydrodynamics and we do this by letting $E\propto t^{\delta _{E}}$ and
$n\propto t^{\delta_{n}}$ where $t$ is the time in the frame of the
observer; we note that the standard model has $\delta_E=\delta_n=0.$
Using (\ref{eq:standard1}) and (\ref{eq:standard2}), we find:
\begin{equation}
 \alpha (\nu>\nu_c)=(p+2)\delta _{E}/4-3p/4+2/4
 \label{eq:idrodinamica1} 
\end{equation}
 and
\begin{equation}
 \alpha (\nu<\nu_c)=(p+3)\delta _{E}/4+\delta_{n}/2-3p/4+3/4
 \label{eq:idrodinamica2}
\end{equation}
Substituting $\alpha_{\rm X}=-1.46$ and $\alpha_{\rm opt}=-1.1$ into (\ref{eq:idrodinamica1}) and (\ref{eq:idrodinamica2}) respectively, we obtain $\delta_E\cong-0.39$, so the shock should lose energy, and $\delta_n\cong0.41$. With those values for $\delta_{E}$ and $\delta_{n}$ we obtain $R(t)\propto t^{0.05}$, where $R$ is the radius of the shock. This result
follows from the fact that $R(t)\sim (Et/n)^{1/4}$. Thus, the density should increase rapidly with radius, as $n\propto R^8$. 

An increase in the density could let one think to the termination shock radius of the wind model by \citet{Chevalier1999} or to the GRB jet running into a sharp-edged dense cloud.
Indeed, \citet{Vietri1999} suggest that the X-ray flare in
GRB 970508 and GRB 970828 could arise from thermal bremsstrahlung
emission of a shock-heated thick torus surrounding the central source.
However, the observed X-ray spectrum is too steep to be compatible with
this model. \citet{Shi1999} invoke a cloudlet to explain the
2-day radio flare \citep{Kulkarni1999b}. 
The energy demands of this model are of the order of $10^{55}$ ergs, but \citet{Sari1999a} have proposed a simpler model in which the radio flare arises in the same reverse shock which powers the prompt optical emission. Moreover, in addition to the necessity of justifying such a density profile, we also need to consider which observable effects could be expected from it e.g., the optical-near IR broad-band spectrum of the burst becoming redder with time if the medium encountered by the fireball becomes very dense. \citet{Holland2000} have shown that there is no evidence for $\beta_{\rm opt}$ varying with time between about $6$ hr and $3$ days since the burst, so that no observable reddening effect seems to become important during the afterglow observations. Using the relations $n_{1}\propto t^{0.41}$ and $R\propto t^{0.05}$, we can estimate that during the afterglow phase, while expanding for about $R(3~{\rm d})-R(6~{\rm hr})=\left[(3~{\rm d}/6~{\rm hr})^{0.05}-1\right]R(6~{\rm hr})\cong 0.13~R(6~{\rm hr})\cong 10^{16}$ cm in radius, the fireball should encounter a medium whose density increases of a factor $(3~{\rm d}/6~{\rm hr})^{0.41}\cong 3$; this means that starting with a typical value of $n_{1}=1$, the afterglow phase will entirely develop while the fireball is expanding in a region where the ISM number density is in the range of $1-3$ particles/cm$^{3}$. So, if we think the $n\propto R^{8}$ profile only limited to the region of space where it is required to explain the afterglow observations, no extreme values of the ISM number density would necessarily be implied by this model, with no important reddening effects expected to be observed.  

The simplest way to lose energy is to assume that the shock is
radiative. However, from the optical to X-ray normalization we know that $\nu_c\cong 1{\rm \rm keV}\gg\nu_m$ and thus the shock is not radiative. A possible mechanism of energy loss was recently suggested in relation with the issue of accounting for bumps observed in the light curves of some GRBs \citep[e.g. ][]{Schaefer2003,Klose2004}; \citet{Schaefer2003} proposed that the bumps observed in the optical transient of GRB 021004 could be related to inhomogeneities in the external gas in the form of clumps of denser material \citep{Wang2000} which should increase the afterglow brightness by enhancing the dissipation of the kinetic energy in the GRB remnant. This process could be of interest in constructing a physical framework for this alternative scenario, since we find the energy loss being connected with an increase in the density. Another possibility is energy loss via cosmic
rays \citep[e.g. ][and references therein]{Waxman1995,Wick2004}. We know little about the energy carried off in
cosmic rays in such strong shocks and so this possibility
requires further inspection. 
 
\section{Conclusions}
We have reported on \textit{Beppo}SAX observations of GRB~990123 and discussed them in the frame of the standard fireball model. We analyzed the broad-band spectrum of the prompt emission, confirming the suggestion of a reverse shock origin for the optical flash.
 
We have studied the properties of the $0.1-60$~keV spectrum and compared the X-ray afterglow with the optical observations and the $8.46$~GHz upper limits. The temporal slopes of the $2-10$~keV and optical light curves are readily explained assuming $\nu_{c}$ between the optical and the X-ray energy band during the first $20$ ks of the \textit{Beppo}SAX follow-up observations, and setting $p=2.47$ for the index of the electron energy distribution; this implies $\beta_{\rm X}=-1.23$. Fitting the $0.1-60$~keV spectrum with a single power-law model, yields $\beta_{\rm X}=-0.82\pm0.10$, which is flatter than the value of $\beta_{\rm X}=-1.23$ expected from the temporal slopes. 

We tried to relate the presence of this  X-ray excess to the contribution of IC scattering and studied a model of synchrotron plus IC emission. By constraining $F^{\rm syn}_{2 \rm keV}$, $F_{m}^{\rm IC}$, $\nu_{c}$ and $\nu^{\rm IC}_{m}$ at the time of the NFIs observations, we have found the corresponding values of the intrinsic parameters of the fireball model: with $\epsilon_{e}\cong10^{-1}$, $\epsilon_{B}\cong10^{-10}$, $E_{52}\cong7\times10^{4}$ and $n_{1}\cong10^{2}$, this model explains both the temporal and spectral behavior of the optical and X-ray afterglow, but violates the $8.46$~GHz upper limits. 

We have compared and discussed our choice for the values of the intrinsic parameters, underlining the importance of the $0.1-60$~keV data as a new piece of information that should be taken into account when fitting the radio, optical and $2-10$ ~keV light curves. 

Finally we have proposed an alternative scenario, where the problem of interpreting the observed value of both the spectral and temporal indices in the optical and X-ray energy band is solved by leaving unchanged the emission mechanism (only synchrotron one) but modifying the hydrodynamics: in this framework, the totality of the observations could only be explained by invoking an ambient medium whose density rises rapidly with radius and by having the shock losing energy.  

\begin{acknowledgements}
The \textit{Beppo}SAX satellite was a joint program of Italian (ASI) and Dutch (NIVR) space agencies. The authors thank S. R. Kulkarni and R. Sari for contributing to an early version of this paper. L. Piro, E. Costa and M. Feroci acknowledge the support of the EU through the EU FPC5 RTN ``Gamma-ray burst, an enigma and a tool''. 
\end{acknowledgements}

\appendix
\section{Details of the Synchrotron+IC Model}
The standard fireball theory for GRBs contains five parameters: the index $p$ of the power-law electron energy distribution; the ratio $\epsilon_{e}$ of the energy in electrons to the post-shock energy in nucleons; the ratio $\epsilon_{B}$ of the magnetic field energy density to the post-shock nucleon energy density; the initial blast wave energy $E_{52}$ (in units of $10^{52}$~ergs) and the ambient number density $n_{1}$ (in units of particles/cm$^{3}$). The value of the index $p$ is generally estimated observing the decay slopes of the afterglow light curves or spectra. To find the values for the other four parameters, it is necessary to express them as functions of observable quantities, such as the characteristic frequencies of the GRB spectrum or the amplitude of the observed flux at a given frequency. \citet{Wijers1999} find the expressions for the intrinsic parameters of the fireball model as functions of the self-absorption frequency $\nu_{a}$, of the peak frequency $\nu_{m}$, of the cooling frequency $\nu_{c}$ and of the peak flux $F^{\rm syn}_{m}$ of the synchrotron spectrum. \citet{Sari2001} find these expressions having also added the effect of IC scattering to synchrotron emission.

In the case of GRB~990123 (see section 3.2 of this paper) 
we have estimates for $\nu_{c}$, $F^{\rm syn}_{2\rm keV}$, $\nu^{\rm IC}_{m}$ and $F^{\rm IC}_{m}$. Thus, we need to express the intrinsic parameters $\epsilon_{e}$, $\epsilon_{B}$, $E_{52}$ and $n_{1}$ as functions of those quantities. We start writing the direct relations that express $\nu_{c}$, $F^{\rm syn}_{2\rm keV}$, $\nu^{\rm IC}_{m}$ and $F^{\rm IC}_{m}$ as functions of $\epsilon_{e}$, $\epsilon_{B}$, $E_{52}$ and $n_{1}$ and then we invert them.

For the peak frequency and the peak flux of the synchrotron spectrum is \citep{Granot1999a}:
\begin{equation}
\nu_{m}({\rm Hz})=(2.9\times10^{15})(1+z)^{1/2}\left(\frac{p-2}{p-1}\right)^{2}\epsilon_{B}^{1/2}\epsilon_{e}^{2}E^{1/2}_{52}t_{d}^{-3/2}
\label{eq:A1}
\end{equation}
\begin{equation}
F^{\rm syn}_{m}(\mu{\rm Jy})=(1.7\times10^{4})(1+z)\epsilon_{B}^{1/2}E_{52}n_{1}^{1/2}d_{28}^{-2}
\label{eq:A2}
\end{equation}
where we have set $\phi_{\rm peak}(p)/\phi_{\rm peak}(2.5)\cong1$ and $\psi_{\rm peak}(p)/\psi_{\rm peak}(2.5)\cong1$ in equation (25) and (26) of \citet{Granot1999a}, because those are slowly varying functions of $p$ \citep{Granot1999a} and we are interested to the case of $p=2.47$. Note that (\ref{eq:A1}) is equivalent to equation (25) of \citet{Granot1999a} when setting $f(p)=\left(\frac{p-2}{p-1}\right)^{2}$ and normalizing for $f(2.5)$ i.e., $2.9\times10^{15}(1+z)^{1/2} f(2.5)\left(\frac{p-2}{p-1}\right)^{2}\frac{1}{f(2.5)}\epsilon_{B}^{1/2}\epsilon_{e}^{2}E^{1/2}_{52}t_{d}^{-3/2}=3.22\times10^{14}(1+z)^{1/2}\frac{f(p)}{f(2.5)}\epsilon_{B}^{1/2}\epsilon_{e}^{2}E^{1/2}_{52}t_{d}^{-3/2}$ \citep{Granot1999a}.

In the case of slow cooling IC-dominated phase, for the synchrotron cooling frequency is \citep{Sari1998,Sari2001}:
\begin{equation}
\nu_{c}({\rm Hz})=(2.7\times10^{12})\epsilon_{B}^{-3/2}E_{52}^{-1/2}n_{1}^{-1}t_{d}^{-1/2}(1+z)^{-1/2}x^{-2}
\label{eq:A3}
\end{equation}
where $x$ is the ratio of the the IC luminosity to the synchrotron one \citep{Sari2001}:
\begin{equation}
x\cong\sqrt{\frac{\epsilon_{e}}{\epsilon_{B}}}\left(\frac{t}{t_{0}^{\rm IC}}\right)^{\frac{p-2}{2(p-4)}}
\label{eq:A4}
\end{equation}
and $t^{\rm IC}_{0}$ is the time at which the slow cooling phase starts \citep{Sari2001}:
\begin{equation}
t^{\rm IC}_{0}(d)=170(1+z)\epsilon_{e}^{3}\epsilon_{B}E_{52}n_{1}\label{eq:A5}
\end{equation}
Note that (\ref{eq:A3}) is equivalent to equation (4.4) of \citet{Sari2001} when expressing $\epsilon_{B}$ in units of $10^{-2}$ i.e., $(2.7\times10^{12})(10^{-2})^{-3/2}\left(\frac{\epsilon_{B}}{10^{-2}}\right)^{-3/2}E_{52}^{-1/2}n_{1}^{-1}t_{d}^{-1/2}(1+z)^{-1/2}x^{-2}=2.7\times10^{15}\epsilon_{B, -2}^{-3/2}E_{52}^{-1/2}n_{1}^{-1}t_{d}^{-1/2}(1+z)^{-1/2}x^{-2}$ \citep{Sari2001}.

The peak frequency of the IC component is \citep{Sari2001}: 
\begin{equation}
\nu_{m}^{\rm IC}=2\gamma_{m}^{2}\nu_{m}
\label{eq:A6}
\end{equation}
where $\gamma_{m}$ is the minimum Lorentz factor of the electrons accelerated in the shock \citep{Sari2001}:
\begin{equation}
\gamma_{m}=(1.116\times10^{4})\epsilon_{e}\left(\frac{p-2}{p-1}\right)\left(\frac{E_{52}}{n_{1}}\right)^{1/8}t_{d}^{-3/8}
\label{eq:A7}
\end{equation}
Note that (\ref{eq:A7}) is equivalent to equation (4.6) in  \citet{Sari2001} when expressing $\epsilon_{e}$ in units of $0.5$ and calculating $\left(\frac{p-2}{p-1}\right)$ for $p=2.2$, i.e. $1.116\times10^{4}\times0.5\left(\frac{\epsilon_{e}}{0.5}\right)\left(\frac{2.2-2}{2.2-1}\right)\left(\frac{E_{52}}{n_{1}}\right)^{1/8}t_{d}^{-3/8}=930~\epsilon_{e, 0.5}\left(\frac{E_{52}}{n_{1}}\right)^{1/8}t_{d}^{-3/8}$ \citep{Sari2001}.

The peak flux of the IC component is \citep{Sari2001}:
\begin{equation}
F_{m}^{\rm IC}(\mu{\rm Jy})=(2\times10^{-7})n_{1}R_{18}F^{\rm syn}_{m}(\mu Jy)
\label{eq:A8}
\end{equation}
where $R_{18}$ is the distance in units of $10^{18}$~cm \citep{Sari1998,Sari2001}:
\begin{equation}
R_{18}=0.39\left(\frac{E_{52}}{n_{1}}\right)^{1/4}t_{d}^{1/4}
\label{eq:A9}
\end{equation}

In these formulas $t_{d}$ is the time measured since the burst in the observer's frame, in units of days, and $d_{28}$ is the luminosity distance of the GRB source in units of $10^{28}$~cm.

Further, if we are in the slow cooling phase, the synchrotron flux at a frequency $\nu>\nu_{c}$ is \citep{Sari1998}:
\begin{equation}
F_{\nu}=F^{\rm syn}_{m}\left(\frac{\nu_{c}}{\nu_{m}}\right)^{\frac{-(p-1)}{2}}\left(\frac{\nu}{\nu_{c}}\right)^{-\frac{p}{2}}
\label{eq:A10}
\end{equation}

We have thus collected all the quantities necessary to calculate the direct relations that express $\nu_{c}$, $F_{\nu}^{\rm syn}$, $\nu_{m}^{\rm IC}$ and $F_{m}^{\rm IC}$ as functions of $\epsilon_{e}$, $\epsilon_{b}$, $E_{52}$ and $n_{1}$.
Substituting (\ref{eq:A5}) in (\ref{eq:A4}) and using the result in (\ref{eq:A3}) we get the following expression for the synchrotron cooling frequency:
\begin{equation}
\nu_{c}({\rm Hz})=A_{\nu_{c}}\left(\epsilon_{B}E_{52}\right)^{a}n_{1}^{b}\epsilon_{e}^{c}
\label{eq:invt_1}
\end{equation}
where:
\begin{equation}
A_{\nu_{c}}({\rm Hz})=(2.7\times10^{12})(1+z)^{\frac{p}{2(p-4)}}(170)^{\frac{p-2}{p-4}}t_{d}^{\frac{8-3p}{2(p-4)}}
\end{equation}
\begin{equation}
a=\frac{p}{2(p-4)}
\end{equation}
\begin{equation}
b=\frac{2}{p-4}
\end{equation}
\begin{equation}
c=2\left(\frac{p-1}{p-4}\right)
\end{equation}

Using (\ref{eq:A1}) and (\ref{eq:A7}) in (\ref{eq:A6}), we obtain for the peak frequency of the IC component:
\begin{equation}
\label{eq:invt_2}\nu^{\rm IC}_{m}({\rm Hz})=A_{\nu^{\rm IC}_{m}}\epsilon_{B}^{1/2}\epsilon_{e}^{4}E_{52}^{3/4}n_{1}^{-1/4}
\end{equation}
where:
\begin{equation}
\nonumber A_{\nu_{m}^{\rm IC}}({\rm Hz})=2(2.9\times10^{15})(1.1\times10^{4})^{2}(1+z)^{1/2}\left(\frac{p-2}{p-1}\right)^{4}t_{d}^{-9/4}
\end{equation}

 Substituting (\ref{eq:A9}) and (\ref{eq:A2}) in (\ref{eq:A8}) we get the expression for the peak flux of the IC component:
\begin{equation}
F_{m}^{\rm IC}(\mu{\rm Jy})=A_{F_{m}^{\rm IC}}\left(E_{52}n_{1}\right)^{5/4}\epsilon_{B}^{1/2}
\label{eq:invt_3}
\end{equation}
where:
\begin{equation}
A_{F_{m}^{\rm IC}}(\mu{\rm Jy})=0.39(2\times10^{-7})(1.7\times10^{4})(1+z)d_{28}^{-2}t_{d}^{1/4}
\end{equation}

Finally, using equations (\ref{eq:A1}), (\ref{eq:A2}) and (\ref{eq:A3}) in (\ref{eq:A10}), we get for the synchrotron flux at a frequency $\nu>\nu_{c}$:
\begin{equation}
F_{\nu}(\mu{\rm Jy})=A_{F_{\nu}}\epsilon_{B}^{d}n_{1}^{e}E_{52}^{f}\epsilon_{e}^{g}
\label{eq:invt_4}
\end{equation}
where:
\begin{equation}
A_{F_{\nu}}(\mu{\rm Jy})=(1.7\times10^{4})(2.7\times10^{12})^{1/2}(2.9\times10^{15})^{\frac{p-1}{2}}\nu({\rm Hz})^{-p/2} d_{28}^{-2}\left(\frac{p-2}{p-1}\right)^{p-1}(170)^{\frac{p-2}{2(p-4)}}(1+z)^{\frac{p^{2}-12}{4(p-4)}}t_{d}^{\frac{-3p^{2}+12p-4}{4(p-4)}}\end{equation}
\begin{equation}
d=\frac{p^{2}-2p-4}{4(p-4)}
\end{equation}
\begin{equation}
e=\frac{p-2}{2(p-4)}
\end{equation}
\begin{equation}
f=\frac{p^{2}-12}{4(p-4)}
\end{equation}
\begin{equation}
g=\frac{p^{2}-4p+3}{p-4}
\end{equation}

The next step is to invert (\ref{eq:invt_1}), (\ref{eq:invt_2}), (\ref{eq:invt_3}), (\ref{eq:invt_4}) and find the expressions for $\epsilon_{B}$, $\epsilon_{e}$, $E_{52}$ and $n_{1}$ as functions of $p$, $\nu_{c}$, $\nu^{\rm IC}_{m}$, $F^{\rm syn}_{\nu}$ and $F_{m}^{\rm IC}$, the value of which we can constrain by using the observed spectral and temporal features of the burst afterglow.

The results of our calculations are the following:
\begin{equation}
\epsilon_{B}=A_{\nu_{c}}^{-a_{1}}A_{\nu_{m}^{\rm IC}}^{-a_{2}}A_{F_{m}^{\rm IC}}^{-a_{3}}A_{F_{\nu}}^{-a_{4}}\left(\nu_{c}\right)^{a_{1}}\left(\nu_{m}^{\rm IC}\right)^{a_{2}}\left(F_{m}^{\rm IC}\right)^{a_{3}}\left(F_{\nu}\right)^{a_{4}}
\label{eq:1}
\end{equation}
where:
\begin{equation}
a_{1}=\frac{10p^{2}-95p+220}{5p^{2}-19p-4}
\end{equation}
\begin{equation}
a_{2}=\frac{25(-p^{2}+5p-4)}{2(5p^{2}-19p-4)}
\end{equation}
\begin{equation}
a_{3}=\frac{-5p^{2}-31p+204}{2(5p^{2}-19p-4)}
\end{equation}
\begin{equation}
a_{4}=\frac{30(p-4)}{5p^{2}-19p-4}
\end{equation}
\begin{equation}
\epsilon_{e}=A_{\nu_{c}}^{-a_{5}}A_{\nu_{m}^{\rm IC}}^{-a_{6}}A_{F_{m}^{\rm IC}}^{-a_{7}}A_{F_{\nu}}^{-a_{8}}\left(\nu_{c}\right)^{a_{5}}\left(\nu_{m}^{\rm IC}\right)^{a_{6}}\left(F_{m}^{\rm IC}\right)^{a_{7}}\left(F_{\nu}\right)^{a_{8}}
\label{eq:2}
\end{equation}
where:
\begin{equation}
a_{5}=\frac{p^{3}-8p^{2}+22p-24}{2(5p^{2}-19p-4)}
\end{equation}
\begin{equation}
a_{6}=\frac{4p^{2}-17p+4}{2(5p^{2}-19p-4)}
\end{equation}
\begin{equation}
a_{7}=\frac{3p-12}{2(5p^{2}-19p-4)}
\end{equation}
\begin{equation}
a_{8}=\frac{-p^{2}+2p+8}{5p^{2}-19p-4}
\end{equation}
\begin{equation}
n_{1}=A_{\nu_{c}}^{-a_{9}}A_{\nu_{m}^{\rm IC}}^{-a_{10}}A_{F_{m}^{\rm IC}}^{-a_{11}}A_{F_{\nu}}^{-a_{12}}\left(\nu_{c}\right)^{a_{9}}\left(\nu_{m}^{\rm IC}\right)^{a_{10}}\left(F_{m}^{\rm IC}\right)^{a_{11}}\left(F_{\nu}\right)^{a_{12}}
\label{eq:3}
\end{equation}
where:
\begin{equation}
a_{9}=\frac{2p^{3}-14p^{2}+25p-4}{5p^{2}-19p-4}
\end{equation}
\begin{equation}
a_{10}=\frac{p^{2}-5p+4}{2(5p^{2}-19p-4)}
\end{equation}
\begin{equation}
a_{11}=\frac{5p^{2}-17p-12}{2(5p^{2}-19p-4)}
\end{equation}
\begin{equation}
a_{12}=\frac{-4p^{2}+14p+8}{5p^{2}-19p-4}
\end{equation}
\begin{equation}
E_{52}=A_{\nu_{c}}^{-a_{13}}A_{\nu_{m}^{\rm IC}}^{-a_{14}}A_{F_{m}^{\rm IC}}^{-a_{15}}A_{F_{\nu}}^{-a_{16}}\left(\nu_{c}\right)^{a_{13}}\left(\nu_{m}^{\rm IC}\right)^{a_{14}}\left(F_{m}^{\rm IC}\right)^{a_{15}}\left(F_{\nu}\right)^{a_{16}}
\label{eq:E52}
\end{equation}
\begin{equation}
a_{13}=\frac{-2p^{3}+10p^{2}+13p-84}{5p^{2}-19p-4}
\end{equation}
\begin{equation}
a_{14}=\frac{9p^{2}-45p+36}{2(5p^{2}-19p-4)}
\end{equation}
\begin{equation}
a_{15}=\frac{5p^{2}-p-76}{2(5p^{2}-19p-4)}
\end{equation}
\begin{equation}
a_{16}=\frac{4p^{2}-26p+40}{5p^{2}-19p-4}
\end{equation}

We now write the numerical expressions for (\ref{eq:1}), (\ref{eq:2}), (\ref{eq:3}), (\ref{eq:E52}) by substituting the values: $z=1.6004$, $d_{28}=3.7$, $p=2.47$, $t_{d}=0.37$~d, $\nu=2$~keV:
\begin{equation}
\epsilon_{B}\cong9.7\times10^{51}\left(\nu_{c}\right)^{a_{1}}\left(\nu_{m}^{\rm IC}\right)^{a_{2}}\left(F_{m}^{\rm IC}\right)^{a_{3}}\left(F_{\nu}\right)^{a_{4}}
\end{equation}
where:
\begin{equation}
a_{1}\cong-2.27
\end{equation}
\begin{equation}
a_{2}\cong-1.38
\end{equation}
\begin{equation}
a_{3}\cong-2.37
\end{equation}
\begin{equation}
a_{4}\cong2.25
\end{equation}
\begin{equation}
\epsilon_{e}\cong2.2\times10^{-9}\left(\nu_{c}\right)^{a_{5}}\left(\nu_{m}^{\rm IC}\right)^{a_{6}}\left(F_{m}^{\rm IC}\right)^{a_{7}}\left(F_{\nu}\right)^{a_{8}}\end{equation}
where:
\begin{equation}
a_{5}\cong8.32\times10^{-2}
\end{equation}
\begin{equation}
a_{6}\cong0.332
\end{equation}
\begin{equation}
a_{7}\cong0.112
\end{equation}
\begin{equation}
a_{8}\cong-0.335
\end{equation}
\begin{equation}
n_{1}\cong12\left(\nu_{c}\right)^{a_{9}}\left(\nu_{m}^{\rm IC}\right)^{a_{10}}\left(F_{m}^{\rm IC}\right)^{a_{11}}\left(F_{\nu}\right)^{a_{12}}
\end{equation}
where:
\begin{equation}
a_{9}\cong-0.121
\end{equation}
\begin{equation}
a_{10}\cong5.50\times10^{-2}
\end{equation}
\begin{equation}
a_{11}\cong0.575
\end{equation}
\begin{equation}
a_{12}\cong-0.890
\end{equation}
\begin{equation}
E_{52}\cong1.1\times10^{-19}\left(\nu_{c}\right)^{a_{13}}\left(\nu_{m}^{\rm IC}\right)^{a_{14}}\left(F_{m}^{\rm IC}\right)^{a_{15}}\left(F_{\nu}\right)^{a_{16}}\label{eq:Energia_numerica}
\end{equation}
where:
\begin{equation}
a_{13}\cong1.03
\end{equation}
\begin{equation}
a_{14}\cong0.495
\end{equation}
\begin{equation}
a_{15}\cong1.17
\end{equation}
\begin{equation}
a_{16}\cong-8.99\times10^{-3}
\end{equation}
\bibliographystyle{aa}
\bibliography{2532}
\end{document}